\documentclass[onecolumn,sort&compress,numbers]{els-mrw} 

\usepackage{amsmath,amssymb,amsfonts,amsthm,makeidx,graphicx}
\usepackage{txfonts}
\usepackage{helvet}
\usepackage{comment}
\usepackage{cancel}
\usepackage{subcaption}
\usepackage{booktabs}
\usepackage{siunitx}
\newcommand\ptwiddle[1]{\mathord{\mathop{#1}\limits^{\scriptscriptstyle(\sim)}}}

\begin{document}


\chapter{Beyond the Standard Model contributions to dipole moments}\label{chap1}

\author[1]{Nicola Valori}%
\author[2]{Oscar Vives}%


\address[1]{\orgname{Institut de Física Corpuscular (CSIC-Universitat de València)}, \orgdiv{Departament de Física Teòrica}, \orgaddress{Dr. Moliner 50, E-46100 Burjassot (València), Spain}}
\articletag{Chapter Article tagline: update of previous edition, reprint.}

\maketitle

\begin{abstract}[Abstract]
The goal of this work is to provide a pedagogical introduction to dipole moments and dipole transitions in theories beyond the Standard Model, with a focus on the lepton sector. We emphasize the exceptional sensitivity of dipole observables to new physics, analyzing the most sensitive processes, such as the muon anomalous magnetic moment, the electric dipole moment of the electron, and the branching ratio $\text{BR}(\mu \to e \gamma)$. This review is intended for PhD students, early-career researchers, and anyone entering the field for the first time. It does not aim to provide a complete overview of all models in the literature, but rather to serve as an accessible guide, using simple and representative examples from the most well-studied extensions of the Standard Model.
\end{abstract}

\begin{keywords}
 	Review \sep Anomalous Magnetic Moment
 	\sep Electric Dipole Moment\sep Muon to electron photon \sep New Physics \sep CP violation \sep Flavor changing neutral currents \sep Experimental sensitivity 
\end{keywords}

\section*{Objectives}
\begin{itemize}
\item In this review, we aim to present some of the key ideas related to dipole observables and their power in probing new physics (NP). We discuss why dipole moments are among the most sensitive observables for detecting physics beyond the Standard Model (SM), and identify, based on both experimental and theoretical considerations, the most promising dipole observables for near-future exploration.

\item We emphasize that all dipole observables are deeply interconnected and offer a unified framework for their treatment. General expressions for the dipole matrix at one-loop level are provided, along with a discussion of particularly relevant two-loop contributions, such as the Barr-Zee diagrams, which play an important role in models with small Yukawa couplings.

\item This formalism is then applied to study the main features of dipole contributions in several well-known extensions of the SM: two-Higgs-doublet models (2HDM), supersymmetric models, theories with extra $U(1)$ gauge symmetries, and the Standard Model Effective Field Theory (SMEFT). Our focus is not on identifying the models that yield the largest contributions to dipole observables, but rather on illustrating the diverse types of new physics that can affect them. 

\item This is by no means an exhaustive list of SM extensions contributing to dipole transitions. Readers interested in further details are encouraged to consult the extensive literature and existing reviews. Below, we provide a selection of complementary works organized by topic, which may offer deeper insights.
For a general overview on the physics of lepton dipole moments we refer the reader to the book \cite{Roberts:2009xnh}.
For discussions on the muon's anomalous magnetic moment in and beyond the SM, see \cite{Jegerlehner:2017gek, Jegerlehner:2009ry,Athron:2021iuf, Athron:2025ets}. For electric dipole moments, detailed introductions to new physics contributions can be found in \cite{Commins:1999jh, Pospelov:2005pr, Engel:2013lsa}. Finally, for a general overview of $\mu \rightarrow e$ transitions, see \cite{RevModPhys.73.151, Calibbi:2017uvl, Ardu:2022sbt}.
\end{itemize}

\section{Introduction}\label{intro}

Dipole moments, or dipole transitions, are quantities that characterize how fermions interact with external electromagnetic fields. Unlike the usual electromagnetic current, which describes the renormalizable interactions of the electromagnetic field, dipole moments arise from effective interactions associated with higher-dimensional operators.
These operators are suppressed by the scale of physics associated to the loop and can be sensitive to the effects of heavy virtual particles predicted by many extensions of the SM. As we will discuss below, dipole operators can probe a broad range of physical phenomena, including Lepton Flavor Violation (LFV) and CP violation, making them powerful tools in the search for NP. 

Indeed, dipole interactions have played, and continue to play, a pivotal role in testing the SM of particle physics and exploring physics beyond it across a wide range of energy scales. In the following, we focus on dipole observables in the leptonic sector, which are among the most sensitive probes for physics beyond the SM.

Within the leptonic sector, dipole observables can be categorized into three groups: gyromagnetic ratios (or g-factors), which are both experimentally measured and theoretically predicted with exceptional precision within the SM; electric dipole moments (EDMs), which violate CP symmetry and whose current experimental bounds lie many orders of magnitude above SM expectations; and lepton flavor transitions, which violate lepton flavor conservation and are strictly forbidden in the SM (neglecting neutrino masses), thus offering a clean window into potential NP.

The g-factor is defined as the proportionality constant between the magnetic moment ($\vec{\mu}$) and the spin ($\vec{s}$) of a fermion $f$ with charge $q$ and mass $m$:  $ \vec{\mu}_{} =g \; q/(2m) \, \vec{s} $. The historical significance of the g-factor lies in its early role as a test of quantum theory. Dirac’s relativistic theory of quantum mechanics famously predicted that the g-factor of a free, point-like fermion should be exactly 2. However, in 1948, Julian Schwinger computed the leading-order quantum correction to the electron’s magnetic moment, arising from the exchange of a virtual photon, and found an additional term of $\alpha_{\rm em}/{\pi}$ \cite{PhysRev.73.416}, where $\alpha_{\rm em}$ is the electromagnetic fine structure constant.  This correction explained the small, unexpected 0.12\% excess observed in precision measurements of the electron’s magnetic moment, a discrepancy referred to as the anomalous magnetic moment of the electron  \cite{PhysRev.74.250}. This extraordinary success established Quantum Electrodynamics as the correct theory for describing electromagnetic interactions, and quantum field theory as a general framework for the theory of elementary particles.  It is now conventional to define the anomalous magnetic moment of a fermion $f$ as: $a_{f}=(g_{f}-2)/2$, which quantifies the deviation of $g_{f}$ from the Dirac value of 2. Since then, anomalous magnetic moments have been measured with incredible accuracy, becoming the best measured quantity in the SM. Nowadays, the electron's g-factor is known with an accuracy of 0.13 ppb \cite{PhysRevLett.130.071801}, and the muon's g factor with an accuracy of 0.19 ppm \cite{Muong-2:2023cdq}. These measurements, along with the corresponding theoretical calculations at high orders, allow to probe the SM and search for extensions of it \footnote{There are potential discrepancies between the SM predictions and the experimental values of $a_e$ and $a_\mu$. For recent determinations of the SM value and discussions, see Refs.~\cite{Aoyama:2020ynm,Boccaletti:2024guq,Morel:2020dww, Parker_2018}.}. Given the exceptional sensitivity to NP and the current unresolved puzzles, as explained in Section~\ref{Subsec:muong-2}, this work focuses mainly on the anomalous magnetic moment of the muon.

Similarly to magnetic moments, parametrizing the interaction between a particle with spin and a magnetic field, EDMs characterize the interaction between a fermion and an external electric field. In the non relativistic theory, they can be parametrized by $\mathcal{H}_{\rm int} = -d_{f} \, \vec{E} \cdot\hat{s}$, where $d_{f}$ is the EDM of the fermion.
Given the fact that the spin $\vec{s}$ is a pseudo-vector and the electric field $\vec{E}$ is a vector, this interaction term violates parity (and more generally CP). The need for new sources of CP violation, beyond those present in the SM, is well motivated by the observed baryon asymmetry of the universe, which cannot be explained within the SM alone.
Theoretically, EDMs are particularly well-suited for CP violating NP searches because their SM contributions are extremely suppressed. Experiments looking for EDMs are sensitive to much larger values than those predicted
by the SM, and would thus be able to unambiguously detect or constraint CP-odd new physics interactions.
Indeed, within the SM, the lepton EDM appears only at four-loop order, predicting, for the electron, $d_{e} \sim 10^{-44} \, \rm e \cdot cm$ \cite{PhysRevD.89.056006} \footnote{However, the dominant contribution to the lepton EDM arises from long-distance effects, which have been evaluated using hadronic effective models \cite{PhysRevLett.125.241802}. These values are presented in Table~\ref{tab:sensitivities}.}. These values are still very far from the corresponding experimental sensitivities, that place upper bounds of $d_{e} < 4.1 \times 10^{-30} \rm e \cdot cm$ at 90 $\%$ confidence level \cite{Roussy_2023} . 

Finally, the third class of dipole observables are lepton flavor-violating processes, such as $\mu \rightarrow e \gamma$, $\tau \rightarrow e \gamma$, and $\tau \rightarrow \mu \gamma$. In the SM with massive neutrinos, these transitions are extremely suppressed, with branching ratios scaling as $(m_\nu / M_W)^4$. Consequently, this results in  $\mathrm{Br}(\mu\rightarrow e \gamma)\sim 10^{-55} \div 10^{-54}$, far beyond the reach of even the most ambitious experimental efforts. As a result, any observation of LFV in dipole transitions would provide a clear and unambiguous signal of physics beyond the SM with new flavor structures.

The experimental status as well as the theoretical predictions in the SM for the leptonic dipole observables are summarized in Table~\ref{tab:sensitivities}.

This work is organized as follow.
In Section~\ref{sec:DipNewPhys}, we define the dipole operator and show how the various dipole observables are related to it. In Section~\ref{sec:DipBSM}, we provide general expressions for the dipole operator at one-loop level, along with selected two loop effects that maybe relevant. These expressions are then applied to illustrative examples of beyond the SM theories. In Section~\ref{sec:Dipobs}, we summarize the impact of such extensions of SM to dipole observables. Finally, in Section~\ref{sec:conclusions}, we present our conclusions.

\begin{table}[ht]
    \centering
    \renewcommand{\arraystretch}{1.5}
    \small
    \begin{tabular}{>{\centering\arraybackslash}m{3.5cm} >{\centering\arraybackslash}m{5cm} >{\centering\arraybackslash}m{5.4cm}}
        \toprule
        \textbf{Observable} & \textbf{Experiment} & \textbf{SM Prediction} \\
        \midrule
        $a_e$ & 
        \num{115965218.059 (13)}  $ \, \times \,10^{-11}$ \cite{PhysRevLett.130.071801} &
        {\scriptsize
        \begin{tabular}[c]{@{}l@{}}
        \num{115965218.0252(95)} \,$\times \,10^{-11}$ [\(\alpha(\mathrm{Rb})\)] \cite{Morel:2020dww} \\
        \num{115965218.161(23)} \,$\times\, 10^{-11}$ [\(\alpha(\mathrm{Cs})\)] \cite{Parker_2018}
        \end{tabular}} \\
        \midrule
        $a_\mu$ & 
        \num{116592059(22)} \,$\times\, 10^{-11}$ \cite{Muong-2:2023cdq} &
        {\scriptsize
        \begin{tabular}[c]{@{}l@{}}
        \num{116591810(43)} \,$\times \,10^{-11}$ (\(e^+e^-\)) \cite{Aoyama:2020ynm} \\
        \num{116592019(38)} \,$\times \,10^{-11}$ (BMW) \cite{Boccaletti:2024guq}
        \end{tabular}} \\
        \midrule
        $|d_e|$ & 
        $< \num{4.1e-30}~\si{e.cm}$ \cite{Roussy_2023} & 
        $\sim \num{1e-39}~\si{e.cm}$ \cite{PhysRevLett.125.241802} \\
        \midrule
        $|d_\mu|$ & 
        $< \num{1.8e-19}~\si{e.cm}$ \cite{Muong-2:2008ebm} & 
        $\sim \num{1e-38}~\si{e.cm}$ \cite{PhysRevLett.125.241802} \\
        \midrule
        $|d_\tau|$ & 
        $\lesssim \num{1e-17}~\si{e.cm}$ \cite{Belle:2021ybo} & 
        $\sim \num{1e-37}~\si{e.cm}$ \cite{PhysRevLett.125.241802} \\
        \midrule
        BR$(\mu \to e \gamma)$ & 
        $< \num{3.1e-13}$ \cite{MEGII:2023ltw} & 
        $\sim 10^{-55} \div 10^{-54}$ \\
        \midrule
        BR$(\tau \to e \gamma)$ & 
        $< \num{3.3e-8}$ \cite{BaBar:2009hkt} & 
        $\sim 10^{-55} \div 10^{-54}$ \\
        \midrule
        BR$(\tau \to \mu \gamma)$ & 
        $< \num{4.4e-8}$ \cite{BaBar:2009hkt} & 
        $\sim 10^{-55} \div 10^{-54}$ \\
        \bottomrule
    \end{tabular}
    \caption{Summary of experimental measurements and SM predictions for dipole-related observables in the lepton sector. The SM values for $a_e$ use both Rubidium and Cesium determinations of $\alpha_{\rm em}$. For $a_\mu$, we show predictions from $e^+e^-$ data and lattice QCD (BMW). All upper bounds are quoted at 90\% C.L.}
    \label{tab:sensitivities}
\end{table}

\section{Dipole moments and the search for new physics}
\label{sec:DipNewPhys}

To understand the role and importance of dipoles in the search for NP,  it is essential to express the relevant observables within the framework of Quantum Field Theory. In the flavor-diagonal sector, the most general Lorentz-invariant form of the fermion-photon amplitude at loop level reads
\begin{eqnarray} 
\label{eq:GenPhoton}
    i\, {\cal M} \:=\: -i\, e\, Q_\ell\, A_\mu(q)\, \overline{u}_{l}(p')\, \Gamma^\mu(q^2)\, u_{l}(p)\,,
\end{eqnarray}
where $\Gamma^\mu$ generalizes the standard tree-level Lorentz vector, $\gamma^\mu$, as follows \footnote{An additional term, the so called anapole moment, is allowed by the Lorentz structure of the amplitude. However it does not represent a well-defined observable \cite{PhysRevD.43.2956}.} \cite{Aebischer:2021uvt},
\begin{eqnarray}
\label{eq:GamMu}
    \Gamma^\mu \:=\: F_1(q^2)\, \gamma^\mu \;+\; F_2(q^2)\, \frac{i\, \sigma^{\mu\nu} 
                     q_\nu}{2m_\ell} \;-\; F_3(q^2)\, \gamma_5\, \sigma^{\mu\nu} 
                     q_\nu \, .
 \end{eqnarray}  
This parametrization holds to all orders in perturbation theory.
The first form factor $F_{1}(q^2)$ is associated to the renormalization of the electromagnetic current, while the other two leads to the definition of $a_{\ell}$ and $d_{\ell}$ as\footnote{From now on, we simply rewrite $f=\ell$ to stress the fact that we refer to lepton quantities.}:
\begin{equation}
    a_{\ell}=F_{2}(0)\,; \quad \quad d_{\ell}/(Q_{\ell} e) = F_{3}(0)\,,
\end{equation}
where we notice that the $a_{\ell}$ is dimensionless quantity while $d_{\ell}$ has a mass dimension of $ \rm Energy^{-1}$. 

In order to predict contributions to dipole observables in extensions of the SM, we define the dipole operator in the low energy effective Lagrangian:
\begin{equation}
  \label{eq:effL}
    {\cal L} \:\supset\: \frac{e\,}{8\, \pi^2}\, C_{ij}\left(\bar\ell_i \sigma_{\mu \nu} P_R \ell_j\right)\, F^{\mu\nu} \:+\: {\rm h.c.}, 
\end{equation}
where $F^{\mu\nu}=\partial^\mu A^\nu-\partial^\nu A^\mu$ is the electromagnetic field strength tensor and the Wilson coefficient $C_{\ell_{j} \ell_{i}}$ is a matrix in flavor space of mass dimension $\rm Energy^{-1}$ that we call the dipole matrix.

Therefore, in terms of the above Wilson coefficients, the NP contribution to the anomalous magnetic moment, $\Delta a_\ell$, and the electric dipole moment, $d_\ell$, are given by
\begin{equation}
  \label{eq:advsC}
    \Delta a_{\ell_{i}} = {-}\frac{m_{\ell_i}}{2 \pi^2 {Q_{\ell}}} \mbox{\rm Re} (C_{ii})\, ,
\qquad \qquad
 d_{\ell_i} = - \frac{e}{4 \pi^2} \mbox{\rm Im} (C_{ii}),
\end{equation} 
where we take the photon momentum entering the vertex.
Similarly, the off-diagonal entries of the dipole matrix describe LFV transitions  \footnote{Note that this is strictly true only for muon decays, where the only available decay channel in the SM is $\mu \rightarrow e \bar{\nu} \nu$. If we consider tau decays, Eq.~(\ref{eq:BRmueg}) must be corrected by a factor ${\rm BR}(\tau \rightarrow \ell \bar{\nu} \nu) \sim 0.176$, to take into account the fact that $\tau$ allows for hadronic decays.}
\begin{eqnarray}
\label{eq:BRmueg}
{\rm BR}(\ell_{j}\to \ell_{i}\gamma)  =  \cfrac{3\,\alpha_{\rm em} }{\pi\,G_F^2\, m_{\ell_{j}}^2}\, \left(|C_{ji}|^2+|C_{ij}|^2\right).
\end{eqnarray}

\begin{figure}[h]
	\centering
\includegraphics[width=5cm]{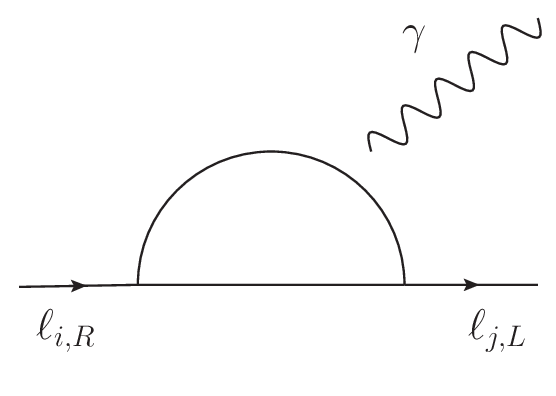}	\caption{Representative topology for a one-loop contribution to the dipole observables. $i,j$ refer to the flavor indices while $R,L$ refer to the chiral component of the external fermions}
	\label{fig:topology}
\end{figure}

At this point, we can understand why dipole transitions are so effective in the search for extensions of the SM. The main reason is that dipole
transitions are induced by a dimension-five operator and must appear at loop level in any renormalizable theory, such as the SM.
This makes them sensitive to heavy virtual particles, such as those predicted by many SM extensions, even if those particles are too massive
to be directly produced in colliders. Figure~\ref{fig:topology} shows a generic topology representing a one-loop contribution to the dipole operator. Internal particles in the loop must be chosen according to the Feynman rules of the theory.

A second relevant aspect of dipole physics is the relationship between dipole and mass operators. Indeed, both of them require a chirality flip, i.e. transitions between left- and right-handed leptons, so that they must be generated by Lagrangian terms that break the chiral symmetry. In the SM, this is carried out only by the Yukawa sector and the Higgs mechanism. This fact guarantees that any contribution to dipole amplitudes is proportional to the SM fermion mass itself. However, due to the mysterious nature of the fermion masses hierarchy, many beyond the SM theories introduce new sources of chiral symmetry breakings, so that contributions to dipole operators may be proportional to the mass of new heavy states. 

In addition to this, if we remove the external photon from the process in Figure~\ref{fig:topology}, we obtain a loop correction to the chirality-changing two-point function. This corresponds to a loop correction to the mass, involving exactly the same interactions and states as the dipole operator. For a pedagogical introduction to the relation between anomalous magnetic moments and masses, see also \cite{Stockinger:2022ata}. If we enhance the dipole contribution by means of heavy fermions in the loop, we expect, by dimensional arguments, that the NP correction to the mass to be of order, $\Delta m_\ell \sim G M_F$ with $G$ the relevant combination of couplings and loop factors from the new interactions and $M_F$ a heavy fermion mass.
Assuming that the loop correction is not larger than the tree-level fermion mass\footnote{This is a naturalness issue. In a strictly technical sense, there is no problem with $\Delta m_\ell > m_\ell$, as we can always cancel this large contribution by an appropriate tuning of the tree-level mass.}, we would expect $G M_F \lesssim \, m_\ell$. Then, this would impose an additional restriction, $\Delta a_\ell \sim G m_\ell M_F/M_F^2 \lesssim m_\ell^2/M_F^2$ and similarly, under certain assumptions, for other dipole observables.  
The interplay between masses and dipoles plays a crucial role in determining the sensitivity of these observables to NP, both from the experimental and theoretical perspectives. 
In particular, anomalous magnetic moments and EDMs are given by the real or imaginary part of the same flavor-diagonal Wilson coefficient, in the basis of real and diagonal charged lepton masses. This implies that contributions to both the dipole matrix and the fermion masses must be taken into account to properly define anomalous magnetic moments and EDMs in the basis of real masses and extract the "observable" dipole moments\footnote{Dipole moments are not strictly basis-independent observables, but since they are extracted from physical processes, they can be embedded in weak basis invariants—typically by combining with chirality-changing operators like fermion masses. See \cite{Botella:2004ks} for a formal treatment.}. 
In the models considered in this review, this issue must be addressed when the loop correction to the mass is of the same order as the tree-level mass, requiring a rephasing of the total mass to extract the observable EDM and magnetic moment values  \cite{Calibbi:2020emz,Calibbi:2021qto}. 

These properties are particularly relevant in the study of anomalous magnetic moments. From an experimental point of view, it is much easier to work with first generation (stable) fermions, making the electron anomalous magnetic moment more precisely measured than the muon magnetic moment, as shown in Table~\ref{tab:sensitivities}. Moreover, as already said, the SM theoretical predictions for these anomalous magnetic moments are known with comparable precision to the experimental measurements in both cases. Therefore, at first sight, it may seem that the electron is more sensitive to NP contributions than the muon. However, this is not always the case. 

Assuming that NP contributions to the anomalous magnetic moment arise at the one-loop level, the chirality change is proportional to the correction to the lepton mass $\Delta m_\ell$. Then, from dimensional analysis and using Eq.~(\ref{eq:advsC}), we find
\begin{equation}
\Delta a_\ell \sim \beta\, \frac{m_\ell^2}{\Lambda_{\rm NP}^2}\, .
\end{equation}

where $\Lambda_{\rm NP}$ is the scale of NP in the loop and $\beta \sim \Delta m_\ell/m_{\ell} $ at most $O(1)$ to respect $\Delta m_\ell \lesssim m_{\ell}$. Thus, assuming a similar $\beta$ factor, the ratio of NP contributions to  $a_\mu$ and $a_e$ is given by
\begin{equation}
\frac{  \Delta a_\mu}{\Delta a_e} \sim \frac{m_\mu^2}{m_e^2} = 4 \times 10^4\,.
\end{equation}
This factor compensates for the difference in the current experimental sensitivities of the muon and electron anomalous magnetic moments. If we take the experimental uncertainties as shown in Table~\ref{tab:sensitivities}, we obtain,
\begin{equation}
  \frac{\delta a^{\rm exp}_\mu}{\delta a^{\rm exp}_e} \sim 10^{3}\, .
  \end{equation}
This implies that, given the current experimental precision, the muon anomalous magnetic moment can probe energy scales roughly one order of magnitude higher than those accessible via the electron.

The situation is quite different for EDMs. Experimental sensitivity is significantly higher for the electron\footnote{The unit e cm is the dipole moment of an $e^+ e^-$ pair separated by 1 cm. The conversion factor into energy units is $\hbar c = 1.9733 \times 10^{-11}$ MeV cm.}, with $|d_e|<4.1\times 10^{-30}$~e cm, compared to $|d_\mu|<1.8 \times 10^{-19}$~e cm. However, by applying an argument similar to that used for the anomalous magnetic moment, NP contributions are proportional to $m_\ell/\Lambda_{\rm NP}^2$, leading to,  
\begin{equation}
\frac{\Delta d_\mu}{\Delta d_e} \sim \frac{m_\mu}{m_e} = 2 \times 10^2\, .
\end{equation}
Therefore, the electron EDM can probe energy scales that are a factor $5 \times 10^4$ larger than the muon EDM.

So, it is clear $d_e$ will remain the main actor in the search for CP violating NP for the foreseeable future, even when accounting for the suppression from the small electron mass.

On the other hand, when exploring off-diagonal dipole transitions, the only viable options are muons or third-generation decays. In this case, muon decays are much better constrained experimentally than tau decays, as shown in Tab.~(\ref{tab:sensitivities}). Therefore, there is no doubt that, in most of the theories beyond the SM, the key processes in flavor off-diagonal transitions will be muon decays, as $\mu \to e \gamma$, $\mu \to e e e $ or $\mu - e$ transitions in nuclei.

\section{Dipole moments beyond the Standard Model}\label{sec:DipBSM}

Despite the tremendous success of the SM in explaining a wide range of phenomena, we remain convinced that NP must exist at high energies to address the various unanswered questions it leaves open. This NP involves new states or interactions that have yet to be detected, either due to experimental limitations or because these states are too heavy to be produced at current facilities.

Nevertheless, any contribution from NP is inherently present, most likely at the loop level, in any measured observable. If NP exists, then by comparing a precisely predicted observable with a sufficiently accurate experimental measurement, we should be able to identify discrepancies that indicate the incompleteness of our theory. Dipole transitions are the best example of this as they are loop processes in the SM. Consequently, NP contributions can compete on equal footing.

NP models are usually based on the introduction of new particles and interactions beyond the SM of particle physics, which, in a renormalizable theory, can be parametrized according to the following Lagrangian:
\begin{align} \label{lagrangian}
    \mathcal{L} \supset - & (y_{R,L})_{ij} \, S \,\bar{\ell}_{i} P_{R,L}F_j\; +\;  (g_{R,L})_{ij}\, \bar{\ell}_{i}\gamma_{\mu} P_{R,L}F_j\, V^{\mu} \;+\; \text{h.c.}\,
\end{align}
where $\ell$ and $A^{\mu}$ are the SM leptons and electromagnetic field, while $F_k$ (fermion), $S$ (scalar) and $V_{\mu}^{(\pm)}$ (massive vector) are left generic, with the only requirement that the interactions preserve the QED gauge symmetry and any other symmetry imposed by the new theory, if present. Interactions between a photon and new particles may also be present, and the Lagrangian term will be just the same as the usual QED.

\begin{figure}[h]
	\centering
	\includegraphics[width=16cm]{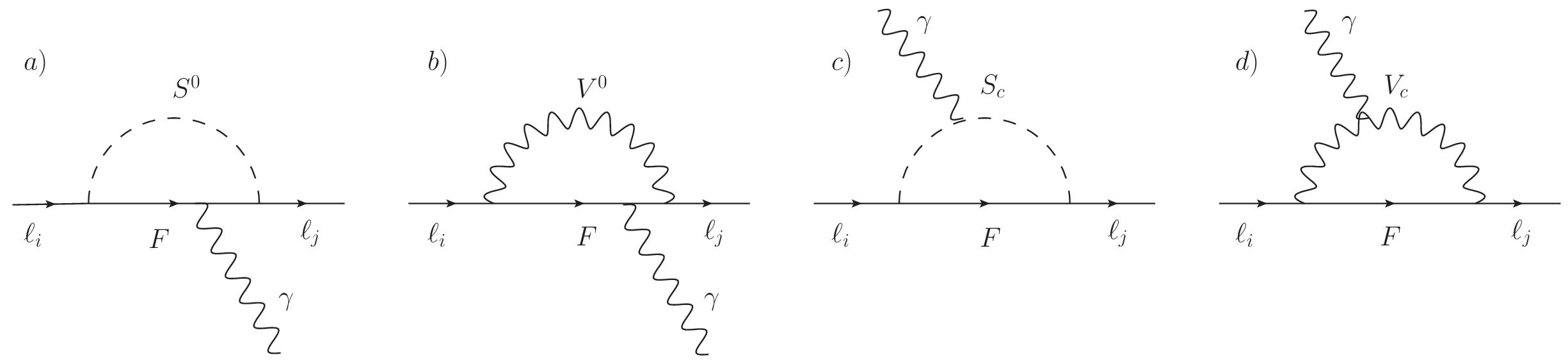}
	\caption{New physics contributions to dipole transitions at one loop: (a) neutral scalar - charged fermion, (b) neutral vector boson - charged fermion, (c) charged scalar - neutral fermion and (d) charged vector boson - neutral fermion.}
	\label{fig:DipoleGen}
\end{figure}

We can make a general classification of NP contributions to dipole transitions at one-loop following Refs.~\cite{Lautrup:1971jf,Leveille:1977rc,Jegerlehner:2009ry}. All the relevant topologies are shown in Figure \ref{fig:DipoleGen}, where we classify them as neutral scalar, Figure~\ref{fig:DipoleGen}(a), neutral vector boson, Figure~\ref{fig:DipoleGen}(b), charged scalar, Figure~\ref{fig:DipoleGen}(c), and charged vector boson, Figure~\ref{fig:DipoleGen}(d).     

We have computed the one-loop contributions to the dipole coefficients from diagrams of Figure~\ref{fig:DipoleGen} in the unitary gauge, where the Goldstone bosons of the theory are absorbed by the massive vector bosons. For a detailed and pedagogical derivation of the dipole diagrams shown in Figure~\ref{fig:DipoleGen}, computed in Feynman gauge and in the context of anomalous magnetic moments, we refer the reader to~\cite{Lynch:2009fxi}. For an example involving diagrams such as Figure~\ref{fig:DipoleGen}(d) in the $R_\xi$ gauge within the context of LFV, see~\cite{Cheng:1984vwu}.

Since the NP particles are usually heavier than the fermions of the first and second generations, i.e. the families most interesting for phenomenology, we present the complete one-loop functions for the dipole coefficients up to order $O(m_{i}/M_{X}^{2})$, where $M_{X}$ is the mass of the vector or scalar boson inside the loop and $m_{i}$ is the mass of the external fermions.
We parametrize the dipole matrix as:
\begin{equation}
\label{eq:DipGen0}
  C_{i j} = Q_{F}C^{(a)}_{i j}  \; + \; Q_{F}C^{(b)}_{i j} \; + \;Q_{s} C^{(c)}_{i j}  \; + \; Q_{V}C^{(d)}_{i j}, 
\end{equation}
where the contributions separately read
\begin{eqnarray} 
  \label{eq:DipGen1}
  C^{(a)}_{i j}&=&\frac{(y_{L})_{ik} (y_{L})_{jk}^* \,m_{i} +  {(y_{R})_{ik} (y_{R})_{jk}^* \, m_{j}}}{ M_s^2 }\; I_1(M_{F_k}^2/M_s^2) \; + \;  \frac{(y_{R})_{ik} (y_{L})_{jk}^* \,M_{F_k}}{M_s^2} \; I_2(M_{F_k}^2/M_s^2) ,\\\label{eq:DipGen2}
C^{(b)}_{i j}&=&\frac{(g_{R})_{ik} (g_{R})_{jk}^* \,m_{i} +  {(g_{L})_{ik} (g_{L})_{jk}^* \, m_{j}}}{ M_V^2 }\; I_3(M_{F_k}^2/M_V^2) \; + \;  \frac{(g_{L})_{ik} (g_{R})_{jk}^*M_{F_k}}{M_V^2} \;  I_4(M_{F_k}^2/M_V^2),
\\\label{eq:DipGen3}
C^{(c)}_{i j}&=&\frac{(y_{L})_{ik} (y_{L})_{jk}^* \,m_{i} +  {(y_{R})_{ik} (y_{R})_{jk}^* \, m_{j}}}{ M_s^2 }\; J_1(M_{F_k}^2/M_s^2) \; + \;  \frac{(y_{R})_{ik} (y_{L})_{jk}^* \,M_{F_k}}{M_s^2} \; J_2(M_{F_k}^2/M_s^2),\\\label{eq:DipGen4}
C^{(d)}_{i j}&=&\frac{(g_{R})_{ik} (g_{R})_{jk}^* \,m_{i} +  {(g_{L})_{ik} (g_{L})_{jk}^* \, m_{j}}}{ M_V^2 }\; J_3(M_{F_k}^2/M_V^2)\; + \; \frac{(g_{L})_{ik} (g_{R})_{jk}^*M_{F_k}}{M_V^2} \; J_4(M_{F_k}^2/M_V^2),
 \end{eqnarray} 
and $Q_i$ are the electromagnetic charges of the particles that couple to the photon in the loop. Then, the loop functions are given by:

\begin{align}
\label{eq:loopfunc}
& I_1(x) =  \frac{1}{48 (1-x)^4} \left[ -2 -3x + 6 x^2 - x^3 - 6 x \log x \right], &
 I_2(x) = \frac{1}{8 (1-x)^3} \left[3 - 4 x + x^2 + 2 \log x \right],
\\
& I_3(x) =  \frac{1}{48 (1-x)^4} \left[8 - 38x + 39 x^2 -14 x^3 + 5 x^4 - 18  x^2 \log x \right], &
  I_4(x) = \frac{1}{8 (1-x)^3} \left[-4 + 3 x + x^3 - 6 x \log x \right],
  \\
&J_1(x) = \frac{1}{48 (1-x)^4} \left[ 1 -6 x + 3 x^2 + 2 x^3 - 6 x^2 \log x \right], &
  J_2(x) = \frac{1}{8 (1-x)^3} \left[1 - x^2 + 2 x \log x \right],
  \\
  &  J_3(x) = \frac{1}{48 (1-x)^4} \left[ -10 + 43 x - 78 x^2 + 49 x^3 - 4 x^4 - 18 x^3 \log x \right], &
J_4(x) = \frac{1}{8 (1-x)^3} \left[4 -15 x +12 x^2 -x^3 - 6 x^2 \log x \right].
\end{align}

The limits of these functions for $x\to 0$, that will be relevant later, are
\begin{align}
\label{eq:loopfunc2}
& I_1(x) \xrightarrow[x\to0]{}  -\frac{1}{24} , &
 I_2(x) \xrightarrow[x\to0]{}   \frac{3}{8} + \frac{\log x}{4},
& &I_3(x) \xrightarrow[x\to0]{}    \frac{1}{6},  &&
  I_4(x) \xrightarrow[x\to0]{}  -\frac{1}{2},
  \\
&J_1(x) \xrightarrow[x\to0]{}   \frac{1}{48}, &
J_2(x)\xrightarrow[x\to0]{}  \frac{1}{8} ,
&  &  J_3(x) \xrightarrow[x\to0]{}  - \frac{5}{24}, &&
J_4(x) \xrightarrow[x\to0]{}  \frac{1}{2}.
\end{align}

Considering the flavor diagonal coefficients, we notice that all the contributions proportional to the external fermion masses are real, and so they only contribute to the anomalous magnetic moment. Then, contribution to EDMs can stem only from the chirally enhanced term, where CP-violation is possible if the theory is chiral (i.e. left-handed SM fermions couple to the NP particles differently from the right-handed ones).

\begin{figure}[h]
	\centering
\includegraphics[width=12cm]{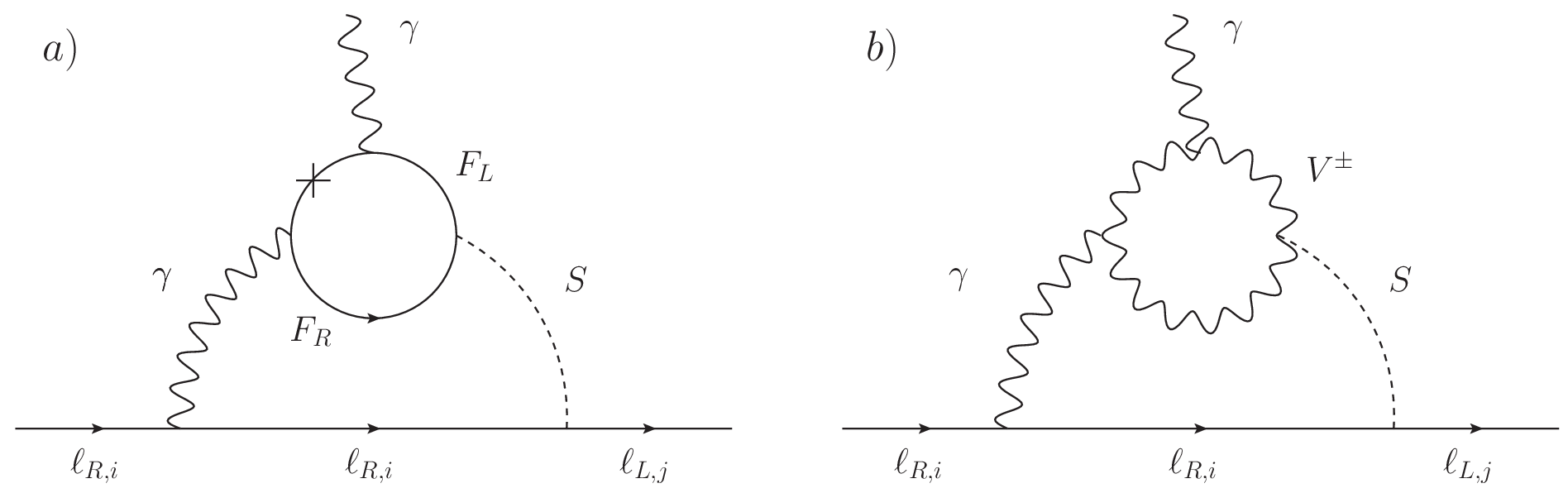}
	\caption{Two-loop “Barr-Zee” diagrams contributing to the dipole transitions with a) charged fermion inner-loop or b) charged vector boson loop.}
	\label{fig:BarrZee}
\end{figure}

In addition, there exist two-loop diagrams, such as the famous Barr-Zee diagrams \cite{Barr:1990vd}, that can be important when one-loop diagrams are suppressed by small couplings, as the first or second generation Yukawa couplings. In these diagrams,  the suppression from the additional loop factor can be compensated by larger couplings.

Figure~\ref{fig:BarrZee} shows two typical Barr-Zee diagrams, where a fermion or boson loop connects to the external fermion line through a (pseudo)scalar and a gauge boson.  For instance, in the context of the SM, these diagrams replace one small Yukawa coupling and a light mass from the diagram in Figure~\ref{fig:DipoleGen}(a) with either two gauge couplings and a larger Yukawa and mass from a heavier fermion ( Figure~\ref{fig:BarrZee}(a)) or three gauge couplings (Figure~\ref{fig:BarrZee}(b)), along with an additional loop factor.  In many beyond the SM theories, the main contribution comes from the diagram where the connecting gauge boson is the SM photon\footnote{Due to Furry's theorem,  only 
 the vector part of the $Z$-$f$-$f$ vertex contributes, and its coupling is accidentally suppressed by a factor $(\frac{1}{4} - \mathrm{sin}^2\theta) \ \sim \ $0.03 compared to the QED one.}. The Barr-Zee contribution to the dipole operator, considering only the dominant photon contribution, is \cite{Chang:1993kw,Ilisie:2015tra}\footnote{For the charged gauge boson expression, we readapted the 2HDM result from Ref.\cite{Ilisie:2015tra}, using $g M_{V}V^{+}V^{-}S$ as the Lagrangian term for the boson-neutral scalar interaction. We considered only the $\xi$-independent part of the computation, since the Barr-Zee diagram from Figure~\ref{fig:BarrZee} b) shows a gauge dependence that cancels out when considering different topologies, as shown in \cite{Abe:2013qla,Altmannshofer_2020}.}:
\begin{align}\label{eq:Barr-Zee}
C_{ij}^{BZ(a)} &=  \frac{\, \alpha\, Q_{F}^2 \, Q_{\ell}}{2\pi \,M_F}\,N_{C}^{F} \,(y_R)_{ij}\,\left[ \text{Re}\left\{ y_F\right\}\,  f(x)\;+\; i\, \text{Im}\left\{ y_F\right\}\, g(x)\,\right], \\
C_{ij}^{BZ(b)} &=- \frac{\alpha \,Q_V^2 \,g \, Q_{\ell}}{8\,\pi\, M_V}\,(y_R)_{ij}\, \left[\,h(x) \right]\,,
\end{align}
where, $x = M^2_{F,V}/M^2_S$, $N_{C}^{F}$ the color factor of $F$ and $y_F$ is the Yukawa couplings between the scalar, $S$, and the internal fermion, $F$,  defined by the Lagrangian term $-y_F\, S \, \bar{F}_{L}F_{R}$. For a pure scalar particle $y_R$ is a real number, while for a pseudoscalar it is purely imaginary.  The loop functions in these expressions are given by
\begin{align}
    f(x) &= \frac{1}{2}x \int_0^1 dz\frac{1-2z(1-z)}{z(1-z)-x} \mathrm{ln}\frac{z(1-z)}{x},\\
    g(x) &= \frac{1}{2}x \int_0^1 dz\frac{1}{z(1-z)-x} \mathrm{ln}\frac{z(1-z)}{x}, \\
    h(x) &= \frac{1}{2}\int_0^1 dz\frac{
x\, z\, \big( 3\, z\, (4 z - 1) + 10 \big) - z\, (1 - z)}{z \,(1 - z) - x
}
\;
\ln\!\left(\frac{z (1 - z)}{x}\right).
\end{align}
As we explained in the previous section, the relationship between dipole and mass operators is very relevant to limit the maximum possible enhancement of dipole operators. Indeed, detaching the external photon line from the diagrams in Figure \ref{fig:DipoleGen} leads to a modification of the lepton mass term in the Lagrangian. 

Though Lagrangian parameters are always scheme dependent and do not represent directly physical observables, the presence of radiative corrections much larger than the measured experimental values necessarily requires a sizable fine-tuning between tree-level parameters and the loop corrections. 
For example, if we consider the scalar loop in Figure~\ref{fig:DipoleGen}(a), the finite correction to SM fermion mass at one-loop in the $\overline{MS}$ scheme is,
\begin{equation} \label{eq:masscorrection}
    (m_{\ell}^{\overline{MS}})_{ij} = m_{\ell}^{0} \delta_{ij}\; - \, (y_R)_{ik}(y_L)^{*}_{jk} \,M_{F_k}\, K_{1}\left(\frac{M_{F_K}^2}{M_s^2}\right), 
\end{equation}
where $m_0$ is the bare mass and the loop function,
\begin{align}
    K_{1}(x)&=\frac{1}{16 \pi^2} \left(\frac{x-1-\mathrm{log}x}{x-1}\right).
\end{align}
From this equation, we see that this expression corresponds to the second term in Eq.(\ref{eq:DipGen1}), with a slightly different loop function. Therefore, a large enhancement in the dipole operators necessarily implies a substantial one-loop correction to the fermion mass. Moreover, these
one-loop corrections to the two point function may induce off-diagonal mass terms or an imaginary contribution to the fermion masses. In order to work in the mass eigenstate basis, the diagonalization and rephasing of the mass matrix must be done order by order before calculating the physical values of anomalous magnetic moments, dipole moments and lepton flavor violating transitions. This procedure can, in turn, induce new flavor or CP violating interactions.

Vector boson loops also contribute to the one-loop corrections to the mass and share the same relation to dipole observables. However, since the fermion mass is not a direct physical observable, the gauge boson contribution is generally gauge-dependent and involves the symmetry-breaking sector. Therefore, we do not discuss it further here.

The expressions presented in this section encompass most of the NP contributions at the one-loop level and highlight the intimate relation between masses and dipole operators. More exotic contributions, such as models with doubly charged fermions or scalars, can be straightforwardly obtained from this by summing the contributions of the charged fermion and charged boson with the corresponding charges. 

In the following sections we will present the contributions to dipole moments from some of the most studied extensions of the SM.

\subsection{Two-Higgs doublet models}

\label{sec:2HDM}

Two Higgs Doublet Models (2HDMs) are the minimal extensions of the SM; they introduce a second Higgs doublet, also associated with the electroweak symmetry-breaking 
scale. The presence of an additional Higgs doublet is motivated by many Beyond the SM scenarios, such as Supersymmetry or Peccei-Quinn models. Moreover, models with multiple scalars allow for new sources of CP violation, both in the scalar potential and the Yukawa sector. For an extensive review on 2HDMs, we refer the reader to \cite{BRANCO20121}.

In these models, we call $\Phi_1$ and $\Phi_2$ the two Higgs doublets, both carrying hypercharge $+1$ \footnote{Here we use the convention where the electric charge is  $Q = T_3 + \frac Y2$}.
\begin{equation}
    \Phi_a = \begin{pmatrix} \phi_a^+ \cr (v_a + \rho_a + i \sigma_a)/\sqrt{2} 
\end{pmatrix}\,.
\end{equation}

We define also the charge-conjugate fields $\tilde \Phi_a\equiv i \tau_2 \Phi_a^*$ with hypercharge $-1$. In general, both scalars acquire vacuum expectation values (vevs), and contribute to the total electroweak symmetry breaking $v =\sqrt{v_1^2+v_2^2} =246$~GeV. We define $\tan \beta = v_2/v_1$, such that the total vev can be written as $v = v_1\cos \beta + v_2 \sin \beta$.

In general, the scalar potential includes several complex parameters which can allow scalar-pseudoscalar mixing. At this point, we neglect this possible CP violating mixing, that we will discuss at the end of the section. Then, the physical spectrum of the 2HDM, after electroweak symmetry breaking, contains a charged Higgs, $H^\pm$, a CP-odd Higgs, $A$, and two CP-even Higgs bosons, $H$  and $h$, which are linear combination of $\rho_1$ and $\rho_2$.
\begin{equation}
\label{eq:CPeigenst}
\begin{pmatrix} h \cr H \end{pmatrix} = \begin{pmatrix} -\sin \alpha & \cos \alpha \cr \cos \alpha & \sin \alpha\end{pmatrix} \begin{pmatrix} \rho_1 \cr \rho_2 \end{pmatrix} .
\end{equation}
Similarly, the physical pseudoscalar, charged Higgs and would-be Goldstone bosons, are given by:
\begin{equation}
\begin{pmatrix} G^{0} \cr A \end{pmatrix} = \begin{pmatrix} \cos \beta & \sin \beta \cr -\sin \beta & \cos \beta\end{pmatrix} \begin{pmatrix} \sigma_1 \cr \sigma_2\end{pmatrix} , \qquad \qquad \begin{pmatrix} G^+ \cr H^+ \end{pmatrix} = \begin{pmatrix} \cos \beta & \sin \beta \cr -\sin \beta & \cos \beta\end{pmatrix} \begin{pmatrix} \phi_1^+ \cr \phi_2^+\end{pmatrix} \,,
\end{equation}
where $G^0 $ and $G^+$ are the would-be Goldstone bosons eaten by the $Z$ and $W$ bosons. 
We can also define the Higgs basis, where only one Higgs, $H_1$, get a vev breaking the electroweak symmetry, that is obtained through the same rotation matrix,
\begin{equation}
\begin{pmatrix} H_1 \cr H_2 \end{pmatrix} = \begin{pmatrix} \cos\beta & \sin \beta \cr -\sin \beta & \cos \beta\end{pmatrix} \begin{pmatrix} \Phi_1 \cr \Phi_2 \end{pmatrix} .
\end{equation}
\begin{table}[b]
	
	{\begin{tabular*}{\textwidth}{@{\extracolsep{\fill}}@{}ccccc@{}}
			\toprule &
			\multicolumn{1}{@{}c}{\TCH{Type I}} &
			\multicolumn{1}{c}{\TCH{Type II}} &
			\multicolumn{1}{c}{\TCH{Type X}} &
			\multicolumn{1}{c@{}}{\TCH{Type Y}}\\
			\colrule
			$\xi^\ell_h$ & $\cos \alpha/\sin \beta $ & $- \sin \alpha/\cos \beta$ & $-\sin \alpha/\cos \beta$ & $\cos \alpha/\sin \beta$ \\[.1cm]
            $\xi^\ell_H$ & $\sin \alpha/\sin \beta$ &  $\cos \alpha/\cos \beta$ & $\cos \alpha/\cos \beta$ & $\sin \alpha/\sin \beta$ \\[.1cm]$\xi^\ell_A$ & $\cot \beta$ & $-\tan \beta$ & $-\tan \beta$ & $\cot \beta$ \\
			\botrule
	\end{tabular*}}{%
    }%
    \caption{Mixing factors in the leptonic Yukawa interactions for natural flavor conservation 2HDM.}\label{tab:xifact}
\end{table}
The most general Yukawa Lagrangian for the leptonic sector in 2HDM is given by,
\begin{align}
\label{eq:Yukawa0}
- \,{\mathcal L}_\text{yukawa}^\text{2HDM} \; = \;
& {\overline
L}_i \ell_{R,j} \, Y^{(1)}_{ij} \,\Phi_1\; + \;  {\overline
L}_i \ell_{R,j} \,Y^{(2)}_{ij}\, \Phi_2\; +\;\text{H.c.}
\end{align}
Using the definitions
above and in the limit of CP conservation, the interactions of the physical scalar mass eigenstates, in the basis of diagonal charged lepton masses, are,
\begin{align}
\label{eq:Yukawa1}
- \,{\mathcal L}_\text{yukawa}^\text{2HDM} = 
&\; {\overline
\ell}_{L,i} \ell_{R,j} \,\frac{\tilde Y_{ij}}{\sqrt2} \, \left(H \cos (\beta- \alpha) + h \sin ( \beta-\alpha ) \right)\; + \; {\overline
\ell}_{L,i} \ell_{R,j} \, \frac{\rho_{ij}}{\sqrt2} 
\,\left(- H \sin (\beta-\alpha) + h \cos ( \beta-\alpha) \right) \nonumber \\ + &\; i \,{\overline
\ell}_{L,i} \ell_{R,j} \, \frac{\rho_{ij}}{\sqrt2}\, A \;+ \;  {\overline
\nu}_{L,i} \ell_{R,j} \, \rho_{ij} \, H^+ + \; \text{H.c.}\nonumber \\
= 
&\; {\overline \ell}_{L,i} \ell_{R,j} \, H \left(\delta_{ij} \, \frac{g\,m^\ell_i}{2\,M_W} \, \cos (\beta- \alpha ) - \frac{\rho_{ij}}{\sqrt2} 
\, \sin (\beta- \alpha ) \right)\; + \; {\overline
\ell}_{L,i} \ell_{R,j} \, h \left(  \delta_{ij} \,\frac{g\,m^\ell_i}{2\,M_W} \, \sin (\beta- \alpha ) +  \frac{ \rho_{ij}}{\sqrt2} 
\, \cos ( \beta - \alpha ) \right) \nonumber \\ + &\; i \,{\overline
\ell}_{L,i} \ell_{R,j} \, \frac{ \rho_{ij}}{\sqrt2}\, A \;+ \;  {\overline
\nu}_{L,i} \ell_{R,j} \, \rho_{ij} \, H^+ + \; \text{H.c.}    
\end{align}
where we defined, $\tilde Y_{ij} = \left( Y^{(1)}_{ij} \cos \beta + Y^{(2)}_{ij} \sin \beta \right)$, and ~ $\rho_{ij} =\left( - Y^{(1)}_{ij} \sin \beta + Y^{(2)}_{ij}\cos\beta \right)$. We have used that $\tilde Y$
is the coupling of $H_1$ to the leptons, and therefore, in this basis, $\tilde Y_{ij}/\sqrt2 = \delta_{ij}\, m^\ell_i /v = \delta_{ij}\, g\, m^\ell_i / (2\, M_W)$. 

Nevertheless, a general 2HDM with generic $\rho_{ij}$ couplings, as happens when both doublets couple simultaneously to the fermions, which is called Type III, has usually too large tree-level Flavour
Changing Neutral Currents (FCNC). This is usually avoided imposing a $Z_2$ symmetry to forbid some of these Yukawa interactions, in the models with Natural Flavour Conservation (NFC). In NFC models the two Higgs doublets are distinguished by their $Z_2$ charge, with $\Phi_1 \to - \Phi_1$ and $\Phi_2 \to  \Phi_2$. Based on the fermion transformations under $Z_2$, we can distinguish different models: 
\begin{itemize}
\item Type I 2HDM: all other SM particles are even under $Z_2$, so that 
all fermions couple only to $\Phi_2$.
\item Type II 2HDM: it enforces $d_{iR} \to -d_{iR}$ and $\ell_{iR} \to -\ell_{iR}$, so that $\Phi_1$ couples only to down-quarks and charged leptons and $\Phi_2$ to up quarks and neutrinos (if right-handed neutrinos are also present).
\item Type X (or lepton-specific): all quarks are even and couple to $\Phi_2$, while the charged leptons are odd and couple to $\Phi_1$.
\item Type Y model: the up-type quarks and charged leptons (even) couple to $\Phi_2$, while the down-type quarks (odd) to $\Phi_1$.
\end{itemize} 
Type III models have the general scalar-fermion interactions in Eq.~(\ref{eq:Yukawa1}), but in models with NFC (types I, II, X and Y) the two Yukawa matrices coupling to a given fermion species can be diagonalized simultaneously and FCNCs are avoided. In these cases, the Yukawa Lagrangian is simplified,
\begin{align}
\label{eq:Yukawa2}
{\mathcal L}_\text{yukawa}^\text{2HDM} =
&-\frac{g\,m^\ell_i}{2\, M_W}\,\xi_h^\ell\,{\overline
\ell}_L\ell_R\, h \;-\; \frac{g\,m^\ell_i}{2\,M_W}\,\xi_H^\ell\, {\overline
\ell}_L \ell_R \, H\;- \;i\,\frac{g\,m^\ell_i}{2\,M_W}\,\xi_A^\ell\,{\overline \ell}_L \ell_R\, A \;-\; \frac{g\, m^\ell_i}{\sqrt{2}\,M_W}\, \xi_A^\ell\,\overline{\nu}_L\ell_R\,H^+\;
+\;\text{H.c.}
\end{align}
The $\xi^a_\ell$ couplings for the four NFC versions of the 2HDM are compiled in Table~\ref{tab:xifact}.

Now, to obtain the contributions of 2HDM to dipole moments, we can apply the general formulas in Eqs.~(\ref{eq:DipGen0}--\ref{eq:DipGen4}). The scalar couplings,  $(y_L^{(a)})_i$, $(y_R^{(a)})_i$, of states with defined CP properties, would be, 
\begin{align}
\label{eq:2HDMCoup}
(y_R^{(h)})_{ij} = (y_L^{(h)})_{ij} = \delta_{ij}\,\frac{g\,m^\ell_i}{2\,M_W}\,\xi_h^\ell  &  \qquad (y_R^{(H)})_{ij} = (y_L^{(H)})_{ij} =  \delta_{ij}\,\frac{g\,m^\ell_i}{2\,M_W}\,\xi_H^\ell \cr  (y_R^{(A)})_{ij} = (y_L^{(A)})_{ji} ^* =  i \,\delta_{ij}\,\frac{g\,m^\ell_i}{2\,M_W}\,\xi_A^\ell & \qquad  (y_R^{(H^+)})_{ij} = (y_L^{(H^+)})_{ij} = \delta_{ij}\,\frac{ g\,m^\ell_i}{\sqrt2\, M_W}\,\xi_A^\ell \,.
\end{align}
Thus, we see that all couplings are flavor conserving and proportional to lepton masses. So, the dipole Wilson coefficients are,
\begin{equation}
\label{eq:W2HDMI}
C_{i j} = \delta_{ij} \,\frac{g^2\,m_i^3}{4\,M_W^2} {Q_{\ell}}\, \left(\sum_{a=h,H}\frac{(\xi^l_a)^2}{M_a^2} \,\left(2\,I_1(m_i^2/M_a^2) + I_2(m_i^2/M_a^2) \right) \; + \; \frac{(\xi^l_A)^2}{M_{A}^2} \, \left(2\,I_1(m_i^2/M_A^2) - I_2(m_i^2/M_A^2) \right) \; + \; \frac{(\xi^l_A)^2}{M_{H^+}^2} \, J_1(m_\nu^2/M_{H^+}^2)\right)\,. 
\end{equation} 
We see that, because of the form of the scalar couplings,  all contributions are necessarily proportional to  $m_{i}^{3}$. Therefore, to have a sizable effect, we need to compensate this suppression either with a large $\xi^l_a$ factor, or with a small scalar mass. 

In Type I and type Y 2HDM, we do not expect a large enhancement from the $\xi^l_i$ couplings, as they are either $O(1)$ or proportional to $\cot \beta$,  that can not be very large due to top quark Yukawa perturbativity and constraints on Higgs masses from colliders \cite{Broggio:2014mna,Haller:2018nnx,BRANCO20121}. On the other hand in Type II and Type X, these couplings are enhanced by $\tan \beta$. In Type II or Type X, the Wilson coefficient would be,
\begin{equation}
\label{eq:W2HDMII}
C_{i j} = \delta_{ij} \,\frac{g^2\,m_i^3}{4\,M_W^2} {Q_{\ell}}\,\left(\frac{\sin^2 \alpha}{M_h^2 \cos^2 \beta} \,\tilde I(m_i^2/M_h^2) \,+\, \frac{\cos^2 \alpha}{M_H^2 \cos^2 \beta} \, \tilde I(m_i^2/M_H^2) \,+\, \frac{\tan^2 \beta}{M_A^2} \, \tilde I^\prime (m_i^2/M_A^2)\; + \; \frac{\,\tan^2\beta}{M_{H^+}^2} \, J_1(m_\nu^2/M_{H^+}^2) \right)\; \,, 
\end{equation} 
where we defined $\tilde I (x) = 2\, I_1(x) + I_2(x)$ and $\tilde I^\prime (x) = 2\, I_1(x) - I_2(x)$. 
Given that the masses involved in NFC models are always the SM lepton masses, we can analyze this contribution in the limit $m_\ell/M_S \to 0$,
\begin{align}
\label{eq:Cm02HDMII}
C_{i j} &\simeq \,\delta_{ij}\,\frac{ g^2\,m_i^3}{4 \,M_W^2} {Q_{\ell}}\,\left( \left(\frac{\sin^2\alpha}{M_h^2\,\cos^2 \beta }  \,+\, \frac{\cos^2 \alpha}{M_H^2\,\cos^2 \beta}\right) \,  \left({\frac{7}{24} + \frac{1}{4}\log \frac{m_{i}^2}{M_H^2}}\right) \,+\, \frac{\tan^2 \beta}{M_A^2} \, \left({-\frac{11}{24} - \frac{1}{4} \log \frac{m_{i}^2}{M_A^2} }\right)\; + \; {\frac{\tan^2 \beta}{48\,M_{H^+}^2} }\right)\; \,.
\end{align} 
For later use, it is useful to express the Higgs mixings in terms of the angle $(\beta- \alpha)$, that parametrizes the rotation between the Higgs basis and the mass basis,
\begin{align}
- \frac{\sin \alpha}{\cos \beta} &= \sin (\beta - \alpha) - \cos(\beta - \alpha) \, \tan \beta \cr
 \frac{\cos \alpha}{\cos \beta} &= \cos (\beta - \alpha) + \sin(\beta - \alpha) \, \tan \beta\,.
\end{align}  
A popular limit in 2HDM, given the absence of experimental signals of extra scalars, is the decoupling limit. The decoupling limit refers to the situation in which one of the two Higgs doublets becomes very heavy and effectively decouples from the low-energy spectrum, leaving behind a light Higgs boson that behaves like the SM Higgs. This limit corresponds to $M_H,M_A,M_{H^+} \gg v$ and 
$\sin (\beta-\alpha) \to 1$.

Finally, general type III 2HDM, has more freedom in the scalar couplings, although the requirement of naturalness would constrain both the $Y^{(1)}_{ij}$ and $Y^{(2)}_{ij}$ matrices to be of the order of the corresponding $i$ or $j$ lepton Yukawas, $m_i/v$ or $m_j/v$. 
The scalar couplings in type III model are,
\begin{align}
\label{eq:2HDM3Coup}
&(y_R^{(h)})_{ij} = (y_L^{(h)})_{ji}^* =   -\delta_{ij}\,\frac{g\,m^\ell_i}{2\,M_W}\,\sin (\beta-\alpha ) -  \frac{\rho_{ij}}{\sqrt2} 
\, \cos ( \beta- \alpha)&  \qquad (y_R^{(H)})_{ij} = (y_L^{(H)})_{ji}^* =   - \delta_{ij}\,\frac{g\,m^\ell_i}{2\,M_W}\,\cos (\beta-\alpha) + \frac{\rho_{ij}}{\sqrt2} 
\, \sin ( \beta-\alpha)\cr  &(y_R^{(A)})_{ij} = (y_L^{(A)})_{ji}^* = - i \frac{\rho_{ij}}{\sqrt2} & \qquad  (y_R^{(H^+)})_{ij} = (y_L^{(H^+)})_{ji}^* = - \rho_{ij} \,.
\end{align}
and the one-loop Wilson coefficient is,
\begin{align}
\label{eq:W2HDMIII}
C_{i j}/{Q_{\ell}} =\, & \delta_{ij} \, \frac{g^2\,m_i^3}{4\,M_W^2}\, F(m_i^2) \;+ \;g\,
\frac{\sin (2(\beta-\alpha))}{4\sqrt2 \,M_W} \,\left[ \rho_{ij} \,\left(  m_i^2 \,G(m_i^2) \,+\,  m_j^2 \,G(m_j^2) \right) \,+\, \rho_{ji}^* \, m_i m_j \left(H(m_i^2) + H(m_j^2) \right)\right] \; + \cr &\left(m_i\, \rho_{ki}^* \rho_{kj} + m_j \,\rho_{ik} \rho_{jk}^* \right) \left(\frac{\cos^2 (\beta-\alpha)}{2 M_h^2 } \, I_1(m_k^2/M_h^2) \,+\, \frac{\sin^2 (\beta-\alpha)}{2 M_H^2 } \,  I_1(m_k^2/M_H^2) \,+\,  \frac{I_1(m_k^2/M_A^2)}{2 M_A^2} \,+\, \frac{J_1(m_\nu^2/M_{H^+}^2)}{M_{H^+}^2}\right)\; + 
\cr
&  m_k\,\rho_{ik} \rho_{kj}  \left(\frac{\cos^2 (\beta -\alpha)}{2 M_h^2 } \, I_2(m_k^2/M_h^2) \,+\, \frac{\sin^2 (\beta-\alpha)}{2 M_H^2 } \,  I_2(m_k^2/M_H^2) \,+\, \frac{I_2(m_k^2/M_A^2)}{2 M_A^2} \right) \,, 
\end{align} 
where we defined the functions:
\begin{align}
\label{eq:2HDMIIIFun}
& F(x^2) \, =\, \frac{\sin^2 (\alpha-\beta)}{M_h^2} \,\tilde I(x^2/M_h^2) \,+\, \frac{\cos^2 (\alpha-\beta)}{M_H^2 } \, \tilde I(x^2/M_H^2)\,,  & \qquad H(x^2) \,=\,\frac{I_1(x^2/M_h^2)}{M_h^2 } \,-\, \frac{ I_1(x^2/M_H^2)}{M_H^2 } \,, \cr & G(x^2) \,=\,
 \frac{I_1(x^2/M_h^2) +I_2(x^2/M_h^2)}{M_h^2 } \,-\, \frac{ I_1(x^2/M_H^2)+ I_2(x^2/M_H^2)}{M_H^2 }
\,.& 
\end{align} 

From this expression we see that some of the scalar couplings are not directly proportional to the masses, but that are instead replaced by the $\rho_{ij}$ matrix.  Moreover, this matrix is not necessarily diagonal in the basis of diagonal charged lepton masses.
This feature may increase the 2HDM contribution in some cases, particularly in lepton flavor violating processes. 

Nevertheless, since both $\tilde Y_{ij}$ and $\rho_{ij}$ are linear combinations of $Y^{(1)}$ and $Y^{(2)}$,  it is clear that  $\rho_{ij}$ can not be generally $O(1)$, but both matrices should be determined by the fermion masses. This idea is captured in the Cheng-Sher ansatz \cite{Cheng:1987rs}, where $\rho_{ij} = \lambda_{ij} \sqrt{2\,m_i\,m_j}/v$, with $\lambda_{ij}\sim O(1)$, and experimental constraints confirm this ansatz for scalar masses around the electroweak scale. 

As a consequence, a double insertion of a scalar interaction in one-loop diagrams can be subdominant compared to a two-loop contribution with larger couplings, especially when the first generations are involved. This is the case of the Barr-Zee diagrams, presented in Section \ref{sec:DipNewPhys}.

To understand the relevance of Barr-Zee diagrams, we can compare the diagram in Figure\ref{fig:DipoleGen}(a) with  Figure\ref{fig:BarrZee} (a), which contains a top quark inside the loop.  From Eq.~(\ref{eq:Barr-Zee}), defining the scalar couplings to the top quark, $\xi_{S}^u g\, m_t P_R/(2 M_W)$ (where $S$ denoting the scalar and $P_R$ the right-handed projector), and in the limit of conserved CP, this contribution would be,
\begin{align}
\label{eq:2HBarrZee}
C_{ij}^{t} &= {\frac{3\, \alpha\, Q_{u}^2}{2\pi \,m_t} Q_{\ell}}\, \frac{g\,m_t}{2\,M_W}\,\left[ \left(\frac{g\,m^\ell_i\,\delta_{ij}}{2\,M_W}\,\sin (\beta-\alpha) +  \frac{\rho_{ij}}{\sqrt2} 
\, \cos(\beta -\alpha)\right)\,\xi_h^u\,  f(m_t^2/M_h^2)\;\right. \cr&+ \;\left.\left(\frac{g\,m^\ell_i\,\delta_{ij}}{2\,M_W}\,\cos (\beta-\alpha) -  \frac{\rho_{ij}}{\sqrt2} 
\, \sin ( \beta-\alpha ) \right)\, \xi_H^u\, f(m_t^2/M_H^2)\;-\;\frac{\rho_{ij}}{\sqrt2} \xi_A^u\,\, g(m_t^2/M_A^2)\,\right] \,.
\end{align}

Assuming $\ell = e$, using that $\xi^u_h = \cos \alpha/\sin \beta$, $\xi^u_H = \sin \alpha/\sin \beta$, $\xi^u_A = \cot \beta$ in all the NFC models, and neglecting ${\cal O}(1)$ differences from the loop functions, the one-loop diagram would be proportional to $m_e (m_e/M_S)^2$, while the second would be proportional to {$m_{e}\,(3\, Q_u^2\alpha/2\pi)$}. Considering the ratio $M_W/m_e\simeq 1.6\times 10^5$, the Barr-Zee diagram dominates the one-loop contribution by a factor of roughly $1\times 10^8\, (M_S/M_W)^2$. 

After discussing the CP-conserving 2HDM, we now turn to the case where CP-violating phases are present in the scalar potential. In this scenario, the CP-odd and CP-even neutral scalars mix, and the physical spectrum is described by a full $3\times 3$ mixing matrix. It is usually enough to consider the limit of small scalar-pseudoscalar mixing. Under the assumption that the pseudoscalar has only a small mixing, $\epsilon$, with both scalar fields, we have,
\begin{equation}
\label{eq:3Hmixing}
\begin{pmatrix} h_1 \cr h_2 \cr h_3 \end{pmatrix} = \begin{pmatrix} -\sin\alpha  &  \cos \alpha & - \epsilon\, \left(\cos(\beta -\alpha ) + \sin(\beta- \alpha )\right)   \cr \cos \alpha  & \sin \alpha & - \epsilon\, \left(\cos(\beta -\alpha) - \sin(\beta -\alpha)\right)  \cr  \epsilon \,\left(\cos \beta - \sin \beta\right) &\epsilon\, \left(\cos \beta +\sin \beta\right)&  1 \end{pmatrix} \begin{pmatrix}\rho_1 \cr \rho_2  \cr A\end{pmatrix} \; + \;{\cal O}(\epsilon^2),
\end{equation}
where $\epsilon$ parametrizes the scalar-pseudoscalar mixing.
From here, we can relate the eigenstates $h_i$ with the states of definite CP, that would be the mass eigenstates if CP is conserved,
\begin{align}
    \label{eq:CPVeigen}
    h_1 &\simeq h - \epsilon \left(\cos(\beta -\alpha) + \sin(\beta-\alpha\right)\, A \cr
    h_2 &\simeq H - \epsilon \left(\cos(\beta -\alpha) - \sin(\beta-\alpha\right)\, A \cr 
    h_3 &\simeq A + \epsilon \left(\cos(\beta -\alpha) + \sin(\beta -\alpha) \right)\, h  +\epsilon \left(\cos(\beta -\alpha) - \sin(\beta -\alpha) \right)\, H\,.
\end{align}
The contributions to the Wilson coefficients in the presence of CP violation can be obtained from Eqs.~(\ref{eq:Yukawa1},\ref{eq:Yukawa2}), using these replacements to obtain the new complex scalar couplings. For instance in NFC models, we obtain,
\begin{align}
\label{eq:2HDMCeps}
&(y_R^{(h_1)})_{ij} = (y_L^{(h_1)})^*_{j i} = \delta_{ij}\,\frac{g\,m^\ell_i}{2\,M_W}\,\left(\xi_h^\ell - i \epsilon \left(\cos(\beta-\alpha) +\sin(\beta-\alpha)\right)\,\xi_A^\ell \right)  \cr  &(y_R^{(h_2)})_{ij} = (y_L^{(h_2)})_{ji}^* =  \delta_{ij}\,\frac{g\,m^\ell_i}{2\,M_W}\,\left(\xi_H^\ell -  i \epsilon \left(\cos(\beta-\alpha) -\sin(\beta-\alpha)\right)\,\xi_A^\ell\right)\cr  &(y_R^{(h_3)})_{ij} = (y_L^{(h_3)})_{ji}^* = \delta_{ij}\,\frac{g\,m^\ell_i}{2\,M_W} \left( i\,\xi_A^\ell + \epsilon  (\xi_h^\ell + \xi_H^\ell)\cos (\beta - \alpha) + \epsilon (\xi_h^\ell - \xi_H^\ell)\sin (\beta - \alpha)\right)  \cr &(y_R^{(H^+)})_{i j} = (y_L^{(H^+)})_{j i}^* = \delta_{i j}\,\frac{ g\,m^\ell_i}{\sqrt2\, M_W}\,\xi_A^\ell \,.
\end{align}
The corresponding Wilson coefficients are easily obtained from, Eqs.~(\ref{eq:DipGen1}, \ref{eq:DipGen3}), by making the sum over the three neutral eigenstates with we appropriate replacements in the couplings and adding the charged Higgs.

The Barr-Zee diagrams can also play an important role in the context of flavor violation and with CP violation in the scalar potential. As shown in Eq.~(\ref{eq:3Hmixing}), in this case we have scalar-pseudoscalar mixing and CP violating couplings to SM fermions that can give rise to Barr-Zee contributions to EDMs of the muon or electron. We can give an expression based on the general Lagrangian in Eq.~(\ref{eq:Yukawa1}), including both CP and flavor mixing. In the case, of small scalar-pseudoscalar mixing, using Eq.~(\ref{eq:CPVeigen}), and neglecting terms $\epsilon \,\xi^u_A = \epsilon \, \cot \beta$, we would have, 
\begin{align}
\label{eq:2HBarrZeeCP}
C_{ij}^{t} &= {\frac{3\, \alpha\, Q_{u}^2}{2\pi \,m_t} Q_{\ell}}\, \frac{g\,m_t}{2\,M_W} \Biggl\{ 
\left( \frac{g\,m^\ell_i\,\delta_{ij}}{2\,M_W} \sin(\beta - \alpha)\, + \,\frac{\rho_{ij}}{\sqrt{2}}\, \cos(\beta - \alpha) \right)\, \xi_{h}^u \, f(x_{t1}) \nonumber \\
&\quad + \, \left( \frac{g\,m^\ell_i\,\delta_{ij}}{2\,M_W}\, \cos(\beta - \alpha) - \frac{\rho_{ij}}{\sqrt{2}} \,\sin(\beta - \alpha) \right) \,\xi_{H}^u \, f(x_{t2}) - \frac{\rho_{ij}}{\sqrt{2}} \, \xi_{A}^u \, g(x_{t3}) \nonumber \\
&\quad + i\, \epsilon \,\frac{\rho_{ij}}{\sqrt{2}} \Bigl[ 
(\cos(\beta - \alpha) + \sin(\beta - \alpha)) \, \xi_{h}^u \, f(x_{t1}) 
+ (\cos(\beta - \alpha) - \sin(\beta - \alpha)) \, \xi_{H}^u \, f(x_{t2}) \nonumber \\
&\quad + \left( \, \cos(\beta - \alpha)(\,\xi_{h}^u \,+ \,\xi_{H}^u\,) + \sin(\beta - \alpha)(\,\xi_{h}^u\, - \,\xi_{H}^u\,) \right)\, f(x_{t3}) 
\Bigr] 
\Biggr\} \,,
\end{align}
where $x_{ti} = m_t^2/M_{h_i}^2$. A detailed analysis of the contributions to $\mu \to e \gamma$ and the electron EDM in the context of the 2HDM can be found in \cite{Altmannshofer_2020,Altmannshofer:2024mbj}.

\subsection{Supersymmetry}
\label{sec:SUSYCij}
A very representative example of possible NP contributing to dipole moments, is given by the Minimal Supersymmetric extension of the Standard Model (MSSM) (for a pedagogical introduction to supersymmetry see \cite{Martin:1997ns}). In the MSSM, any SM chiral fermion  is associated to a new spin-0 complex boson, called sfermion, while the SM gauge bosons and scalar fields\footnote{Note that the cancellation of triangle anomalies require the presence of an additional Higgs doublet, so that MSSM is a Type II Two-Higgs-Doublet Model as defined in the previous section.} are associated to new Weyl fermions, called gauginos or higgsinos, which, after electroweak symmetry breaking, mix into neutralinos, $\chi^0$, and charginos, $\chi^+$.

\begin{figure}[h]
	\centering
\includegraphics[width=6cm]{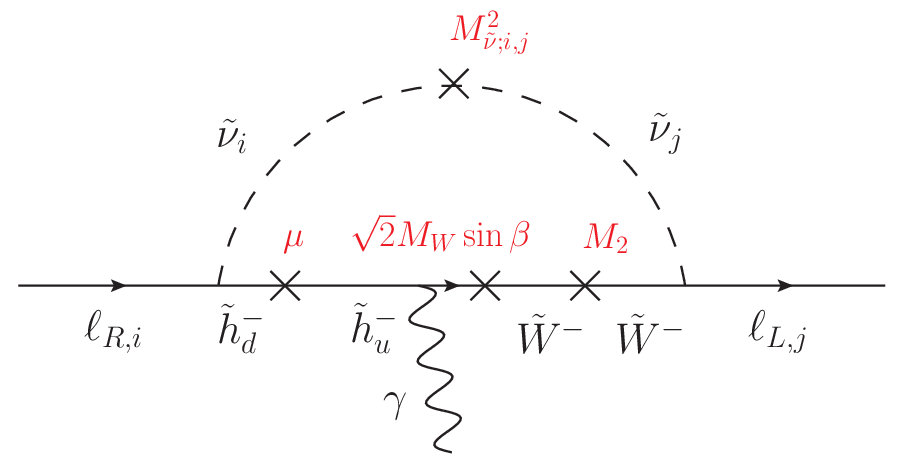}
	\caption{Dominant chargino contribution to dipole transitions, expressed in terms of the higgsino components  $\tilde h_{u}$, $h_d$, and wino, $\tilde W$, within the Mass Insertion Approximation. Mass insertions are denoted by crosses, with their corresponding values indicated in red.}
	\label{fig:Chargino}
\end{figure}

\begin{figure}[h]
	\centering
\includegraphics[width=14cm]{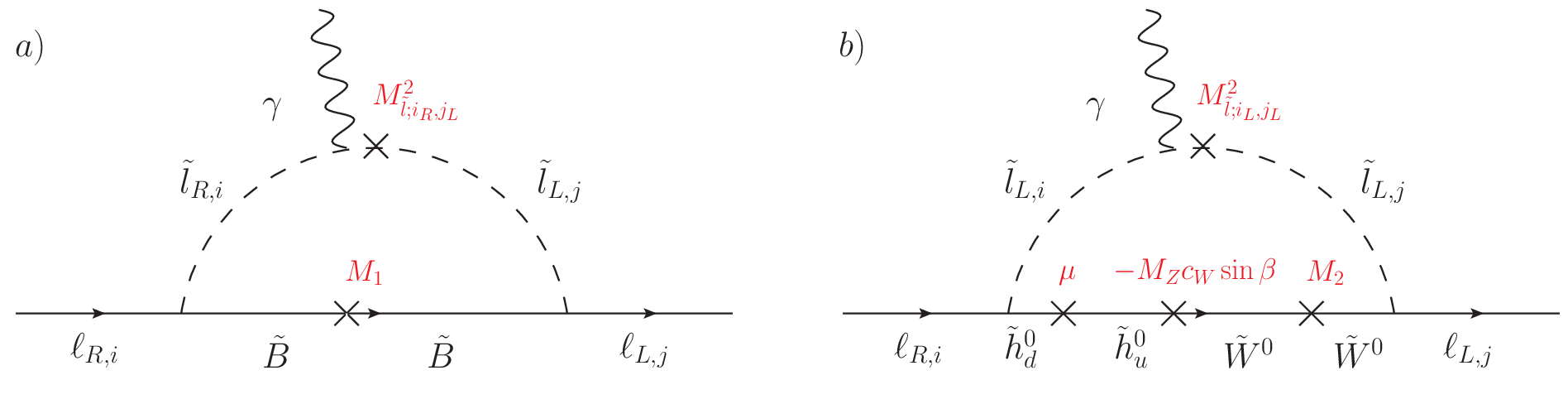}
	\caption{Dominant neutralino contributions to dipole transitions, expressed in terms of the higgsino components  $\tilde h_{u}$, $h_d$, bino, $\tilde B$, and wino, $\tilde W$, within the Mass Insertion Approximation. Mass insertions are denoted by crosses, with their corresponding values indicated in red. Diagram a) represents the pure bino contribution with chirality changing slepton mass insertion and b) is the mixed gaugino-higssino contribution.}
	\label{fig:Neutralino}
\end{figure}

In the MSSM, there are two main contributions to dipole transitions. In the first, the chargino, $\chi^-$, plays the role of the internal fermion in the loop of Figure (\ref{fig:DipoleGen}a), while the sneutrino, $\tilde \nu$ (superpartner of the neutrino), acts as the scalar. 
The second contribution, would correspond to Figure (\ref{fig:DipoleGen}c), where the neutralino, $\chi^0$, is the internal fermion and the charged slepton, $\tilde l$ (scalar superpartner of the lepton), is the scalar \cite{Hisano:1995cp,Hisano:1995nq,Moroi:1995yh} ({see it also in Figures \ref{fig:Chargino} and \ref{fig:Neutralino} below}).

The internal chirality change is given by the chargino or neutralino mass, respectively, providing, at first sight, a large enhancement factor $M_\chi/m_{\ell}$. Chargino masses are constrained to be larger than several hundreds of GeV by LEP and LHC searches, and neutralino are typically a factor two lighter. For instance, in the case of the muon, this factor leads to a $10^3$ enhancement in these diagrams. However, we have to take into account the scalar couplings $y_L$ and $y_R$ , that can be small, as we have seen in the 2HDM, and play an important role in the overall contribution.

We start by analyzing the chargino-sneutrino diagram.

Chargino eigenstates are linear combinations of the fermionic partners of the W-boson, Wino, ($\tilde{W}^{\pm}$), and the charged Higgs, Higgsino $(\tilde{H}^{\pm})$, {which are represented by Weyl bi-spinors. Given the doublets $\psi^{+} =(\tilde{W}^{+},\tilde{H}_u^{+} )^T$, $\psi^{-} =(\tilde{W}^{-},\tilde{H}_d^{-} )^T$ and the mass  Lagrangian term ${(\psi^{-})}^T M \psi^{+}$, the eigenstates are obtained by diagonalizing the mass matrix,}
\begin{equation}
\label{eq:charginomat}
M_{}= \left( \begin{array}{cc} M_2 & \sqrt{2} M_W \sin \beta \\ \sqrt{2}
M_W \cos \beta & \mu  \end{array}\right),
\end{equation} 
{where $\mu$ is the higgsino mass term, $M_2$ the soft SUSY breaking Wino mass and the off-diagonal terms stem from the Yukawa interactions with the Higgs field after the electroweak symmetry breaking.}
{ This matrix is diagonalized with two unitary matrices, $U,V$,}
\begin{equation}
\label{eq:chargdiagon}
 U^* M_{} V^{-1} = \text{Diag.}(M_{\chi_1^\pm}, M_{\chi_2^\pm}).
\end{equation}
Similarly, the left-handed sneutrino mass is diagonalized as,
\begin{equation}
\label{eq:sneutdiagon}
R^{\tilde \nu}  M^2_{\tilde \nu} {R^{\tilde \nu}}^\dagger = \text{Diag.}(M_{\tilde \nu_1},M_{\tilde \nu_2},M_{\tilde \nu_3}).
\end{equation} 
In the end, {given the convention in Eq~(\ref{lagrangian})}, the scalar couplings are given by \cite{Hisano:1995cp},
\begin{eqnarray}
\left({y_R}\right)_{i;a,\alpha}\; =\;  g   {~V_{1a} \; R^{\tilde \nu}_{i \alpha} }\,, \quad & \left({y_L}\right)_{i;a,\alpha} \;= -\;Y_i~  {U_{a2}^*\; R^{\tilde \nu}_{i \alpha}}  \, ,
\end{eqnarray}

where $i$ is the lepton flavor, $a$ labels the chargino eigenstate, $\alpha$ the sneutrino eigenstate. Already at this point, we see that {$y_R$}, as can be seen from the mixing $V_{a,1}$,  comes from the gaugino component and is proportional to the gauge coupling, while  {$y_L$} ({$U_{2a}$} corresponds to higgsino component) involves the lepton Yukawa coupling and is therefore expected to be small.

Form here, the dominant chargino, chirality enhanced, contribution to the dipole Wilson coefficient would be,
\begin{eqnarray}
\label{eq:chargino1}
  C^{\chi^{\pm}}_{i j}\,=\, -{Q_{\chi}}\frac{g^2\, {U_{a2} V_{1a} {R^{\tilde \nu}_{j \alpha} }^* R^{\tilde \nu}_{i\alpha} }}{M_{\tilde \nu}^2}\; \frac{m_i}{\sqrt{2} M_W \cos \beta}   \; M_{\chi^\pm_a}\; I_2(M_{\chi^\pm_a}^2/M_{\tilde \nu_\alpha}^2). 
\end{eqnarray}

In this equation, we have used the fact that the MSSM is a Type II Two-Higgs-Doublet Model, where $H_d$ (and $\tilde h_d$) couples to down-type quarks and charged leptons, while $H_u$ couples to up-type quarks and neutrinos. Then, we have,
\begin{equation}
 Y_i = \frac{g\; m_i}{\sqrt{2}\, M_W \cos \beta} \,.   
\end{equation} 
This coupling appears because of the Higgsino component of the Chargino, indicated by the index $b=2$ in the $U_{a,2}$. On the other hand, the coupling to the left-handed lepton is a much larger gauge coupling, with $V_{1a}$ corresponding to the wino component. 

However,  we can expect a small mixing in the chargino line, going from higgssino to wino state.
This can be seen using the Mass Insertion (MI) approximation \cite{Hall:1985dx,Gabbiani:1996hi}\footnote{
The MI approximation is performed in the interaction basis with diagonal couplings, keeping off-diagonal propagators. The propagator is then expanded in the ratio of (small) off-diagonal to diagonal masses, such that one off-diagonal propagator corresponds to two diagonal propagators plus an off-diagonal mass insertion. The multiple propagators in a loop can be
simplified using the reduction formula: $ 1/((k^2-m_1^2)(k^2-m_2^2)) =  \left[ 1/(k^2-m_1^2) - 1/(k^2-m_2^2)\right] /({m_1^2-m_2^2})$.
This technique helps in tracing back all the mass mixing without diagonalizing the mass matrices and allows to express the result in terms of the loop functions in Eq.~(\ref{eq:DipGen1}-\ref{eq:DipGen4}).
}.

Therefore, from  Eq.~(\ref{eq:charginomat}), following App B. in \cite{Barenboim:2013bla}, for large $\tan \beta$ and in the limit $\mu,M_2 \gg M_W$, we find that,
\begin{eqnarray}
\label{eq:chargino2}
  C^{\chi^\pm}_{i j}\,\simeq -{Q_{\chi}}\,\frac{g^2}{M_{\tilde \nu_\alpha}^2}\; \sqrt{2} M_W\; M_2\, \mu\, \sin \beta\; \frac{m_j}{\sqrt{2} M_W \cos \beta}  \; {R^{\tilde \nu}_{i \alpha}}^* R^{\tilde \nu}_{\alpha j} \frac{I_2(\mu^2/M_{\tilde \nu_\alpha}^2)\, -\,I_2(M_2^2/M_{\tilde \nu_\alpha}^2) }{\mu^2 -  M_2^2} \,,
\end{eqnarray}
Thus, we see that the chiral enhancement $M_{\chi^\pm}/m_i$ is compensated by the scalar couplings, leaving only the well-known $\tan \beta$ enhancement. 

We also apply the MI approximation to the sneutrino line when $i\neq j$, assuming off-diagonal elements are much smaller that diagonal ones which we take all equal to $M_{\tilde \nu}^2$. Then, we have
\begin{eqnarray}
\label{eq:chargino3}
  C^{\chi^\pm}_{i j}\,\simeq\,-{Q_{\chi}} \frac{g^2\,m_j}{M_{\tilde \nu}^4}\;\frac{M_2\,\mu \,\tan \beta}{\mu^2 -  M_2^2}\; \left( \left(M_{\tilde \nu}^2\delta_{ij} \, - \, (M^2_{\tilde \nu})_{ i,j}\right) \,\left(I_2(\mu^2/M_{\tilde \nu}^2)\, -\,I_2(M_2^2/M_{\tilde \nu}^2)\right) \;+\;\frac{(M^2_{\tilde \nu})_{ i,j}}{M^2_{\tilde \nu}} \;  \left(-\mu^2\, I^\prime_2(\mu^2/M_{\tilde \nu}^2)\, + \;M_2^2\, I^\prime_2(M_2^2/M_{\tilde \nu}^2)\right)  \right)  \,,
\end{eqnarray}
with the derivative, $I^\prime_2(x) = d I_2(x)/d x$, obtained in the limit $M_{\tilde \nu_i}^2 =M_{\tilde \nu_j}^2 =M_{\tilde \nu}^2$  with $(M^2_{\tilde \nu})_{ i,i}$= 0. 
We can trace this process in the MI formalism, as illustrated in Figure \ref{fig:Chargino}. In the figure, a right-handed lepton couples to a left-handed sneutrino through the down-type higgsino, $\tilde h_d^-$. 
Along the fermion line, i) the down-type higgsino, $\tilde h_d$, converts into an up-type higgsino, $\tilde h_u$, through a $\mu$ mass term, ii) the up-type higgsino then transforms into a wino through a $\sqrt{2} M_W \sin \beta$ entry in the mass matrix, and iii) to induce a chirality flip, we need another mass insertion, leading to an odd number of insertions, which results in a wino mass,  $M_2$.  In the sneutrino line, we can have flavour change for $i\neq j$ though an off-diagonal element in the sneutrino mass matrix, $M^2_{\tilde \nu; i,j}$. Finally, at the other end of the loop the wino couples the left handed sneutrino with a left handed lepton with a gauge (gaugino) coupling. This is the dominant, $\tan \beta$ enhanced, contribution in Eq.~(\ref{eq:chargino3}).

In a similar way, we can analyze the neutralino contribution. 
In this case, the neutralino mass matrix is symmetric, and, in the basis $(\tilde B, \tilde W^0, \tilde h^0_d,\tilde h^0_u  )$, is given by ~\cite{Haber:1984rc,Martin:1997ns},
\begin{equation}
M_{\chi^0} = \begin{pmatrix}M_1 & 0 & - \cos \beta\, s_W\, M_Z &
\sin \beta\, s_W \, M_Z\cr
0 & M_2 & \cos \beta\, c_W\, M_Z & - \sin \beta\, c_W\, M_Z \cr
-\cos \beta \,s_W\, M_Z & \cos \beta\, c_W\, M_Z & 0 & -\mu \cr
\sin \beta\, s_W\, M_Z & - \sin \beta\, c_W \, M_Z& -\mu & 0 \cr 
\end{pmatrix}\, ,
\label{eq:neutralmatrix} 
\end{equation}
with $s_W = \sin \theta_W$ and $c_W = \cos \theta_W$. This matrix is diagonalized by an orthogonal rotation, 
$N^* M_{\chi^0} N^\dagger = \text{Diag}\left(
M_{\chi^0_1},
M_{\chi^0_2},
M_{\chi^0_3},
M_{\chi^0_4}\right)$.

Then, the $6 \times 6$ slepton mass matrix
\begin{equation}
\label{eq:sleptmass}
M^2_{\tilde l}=\left( \begin{array}{cc}
   M^2_{\tilde l_L} +(s_W^2 -\frac{1}{2})M_Z^2\cos\,  2\beta &
              v_d\; ({Y^A_{\tilde l}}^*- Y_l \;\mu\tan\beta) \\
 v_d\; (Y^A_{\tilde l}-Y_l \;\mu^*\tan\beta) &
    M^2_{\tilde l_R} -s_W^2 \, M_Z^2\cos\,  2\beta 
\end{array}\right),
\end{equation}
with $M^2_{\tilde l_L}$, $M^2_{\tilde l_R}$, $Y_l$ and $Y^A_{\tilde l}$ ({from the sleptons trilinear interaction with the Higgs boson}), all $3\times3$ matrices.
This matrix is diagonalized by 
$R^{\tilde l}  M^2_{\tilde l} {R^{\tilde l}}^\dagger = \text{Diag.}(M_{\tilde l_1},\dots M_{\tilde l_6} )$.

The couplings are now \cite{Hisano:1995cp},
\begin{align}
\left(y_{R}\right)_{i;a,\alpha} &= -\frac{g}{\sqrt{2}} \left\{
       \left[N_{a2} + N_{a1} \tan \theta_W  \right] R^{\tilde l}_{\alpha,i}
        - \;\frac{\sqrt{2}Y_i}{g} \;N_{a3} R^{\tilde l}_{\alpha,i+3} \right\}, \\ \left(y_{{L}}\right)_{i;a,\alpha} &=   -\frac{g}{\sqrt{2}} \left\{ \;-\frac{\sqrt{2}Y_i}{g} \;
           N_{a3} R^{\tilde l}_{\alpha,i} - 2N_{a1} \tan  \theta_W \;R^{\tilde l} _{\alpha,i+3} \right\}, 
\end{align}

{where we used the convention $Q=Y + T_{3}$.} 
From here, we can obtain the chirality-enhanced neutralino contribution as,

 \begin{align}
\label{eq:neutral0}
C^{\chi^0}_{i j}/Q_{\tilde{l}} &= - \frac{g^2}{2  M_{\tilde l_\alpha}^2}   \, \left(2 ~t_{W}^{2} N_{a1}N_{a1}^*  R^{\tilde l}_{\alpha,i}(R^{\tilde l}_{\alpha,j+3})^{*} + \frac{ m_j }{M_W\, \cos \beta} ~t_W \left( N_{a1} N_{a3}^* R^{\tilde l}_{\alpha,i}({R^{\tilde l}_{\alpha,j}})^* - 2\frac{m_i}{m_j} \; N_{a1}^*N_{a3}  ({R^{\tilde l}_{\alpha,i+3}})^*  R^{\tilde l} _{\alpha,j+3} \right)  \right. \cr 
& \left.+ \; N_{a2}R^{\tilde l}_{\alpha,i} \left(2 ~t_W N_{a1}^*  ({R^{\tilde l}_{\alpha,j+3}})^* + \frac{ m_i}{M_W\, \cos \beta} \;
          N_{a3}^* ({R^{\tilde l}_{\alpha,j}})^*\right)
 \right)\; M_{\chi^0_a} \; J_2(M_{\chi^0_a}^2/M_{\tilde l_\alpha}^2)\, ,
 \end{align}

where we neglected the Yukawa squared term.

In this expression, we see that we have three  different terms. Once again, it is useful to apply the MI approximation to track the physics, extracting the odd neutralino mass insertion and expanding the hermitian matrix $M_{\chi^0} M_{\chi^0}^\dagger$, as we did in the chargino case. Following  Ref.~\cite{Barenboim:2013bla}, the final expression for the dominant neutralino contribution is,
\begin{align}
\label{eq:neutral1}
C^{\chi^0}_{i j} / Q_{\tilde{l}}&\simeq - \frac{g^2}{2} \,\mu \tan \beta  \, \left(
 2\, \delta_{ij}\, t_W^2 \frac{m_i\,M_1}{M_{\tilde l_L}^2  M_{\tilde l_R}^2}  \; \frac{ M_{\tilde l_L}^2\,J_2(M_{1}^2/M_{\tilde l_R}^2)\, -\, M_{\tilde l_R}^2\,J_2(M_{1}^2/M_{\tilde l_L}^2) }{(M_{\tilde l_R}^2- M_{\tilde l_L}^2 ) }\, +  t_W^2m_j\,M_1\left[\,\frac{\left((\delta_{ij} M_{\tilde{l}_{L}}^{2}- (M_{\tilde{l}_{L}}^{2})_{ij}\right)}{M_{\tilde{ l}_L}^4} \, \,  \frac{ J_2(\mu^2/M_{\tilde l_L}^2) \,- \,J_2 (M_1^2/M_{\tilde l_L}^2) }{ \mu^2 - M_1^2 }+ \right.\right.\cr 
& \left. \left. +\, \frac{(M_{\tilde l_L}^2)_{ij}}{M_{\tilde l_L}^6}  \; \left(\frac{- \mu^{2}J_{2}^{\prime}(\mu^2/M_{\tilde l_L}^2) +M_{1}^{2}J_{2}^{\prime}(M_1^2/M_{\tilde l_L}^2) }{ \mu^2 - M_1^2 }\right) - 2 \frac{m_i}{m_j}  (L\leftrightarrow R)\right]\,  -\, \frac{m_j\,M_2}{M_{\tilde l_L}^4} \,\left[\left(M_{\tilde l_L}^2\delta_{ij} \, - \, (M^2_{\tilde l_L})_{ i,j}\right) \,\frac{J_2(\mu^2/M_{\tilde l_L}^2)\, -\,J_2(M_2^2/M_{\tilde l}^2)}{\mu^2-M_{2}^{2}} \;+ \right.\right.\cr 
& \left. \left. \;\frac{(M^2_{\tilde l})_{ i,j}}{M^2_{\tilde l}} \;  \frac{\left(-\mu^2\, J^\prime_2(\mu^2/M_{\tilde l}^2)\, + \;M_2^2\, J^\prime_2(M_2^2/M_{\tilde l}^2)\right)}{\mu^2-M_{2}^{2}}\right]\right)\,, 
 \end{align}

where {$(M^2_{\tilde l_{L(R)}})_{ i,j}$ refers only to off-diagonal terms, $i\neq j$, and we take the limit of equal mass for all left-handed and right-handed sleptons, $M_{\tilde l_L}$ and  $M_{\tilde l_R}$}. 

We can see two of these neutralino contributions in Figure \ref{fig:Neutralino}.
The first term, depicted in Figure~\ref{fig:Neutralino}a), has no apparent Yukawa suppression in the neutralino couplings and would correspond to bino exchange. Nevertheless, this term requires a left-right mixing in the slepton line from
$\tilde l_{i_R}$ to $\tilde l_{j_L}$, that from Eq.~(\ref{eq:sleptmass}), in terms of mass insertions, is proportional to $ \delta_{ij}^{LR} = m_i \mu \tan \beta /M_{\tilde l}^2$, where once more, we neglected the term not $\tan \beta$-enhanced taking into account that trilinear couplings are $Y^A_{\tilde l} \sim O( \mu Y_l)$. 

{The second and third terms are similar to the chargino contribution, but with a different loop function. The third term is depicted in Figure~\ref{fig:Neutralino}b), while the second term is diagrammatically the same\footnote{In the bino case, you have also to consider the diagram with $\tilde{l}_{R}$ in the loop.}, except with a $\tilde{B}$ instead of $\tilde{W}^{0}$. In Figure~\ref{fig:Neutralino}b), we go along the fermionic line, from  $\tilde h_d$ to $\tilde h_u$, through a $\mu$  term, then into the wino through a $\sqrt{2} M_W \, \sin \beta$ entry in the mass matrix, and finally we add an $M_2$ for a third chirality flip. On the other hand, the internal slepton is purely left-handed, with a possible flavor change $i \to j$ provided by the mass insertion $(M_{\tilde l_L}^2)_{ij}/M_{\tilde l_L}^2$.}

The full supersymmetric contribution would be the sum of the chargino and neutralino terms, and we present the expressions for the $\tan \beta$ enhanced part.

\subsection{Extra $U(1)$  gauge symmetries}
\label{sec:Zprime}
The addition of a new $U(1)$ gauge symmetry to the SM is one of the simplest ways  to introduce new gauge interactions. The requirement of the new gauge symmetry to be anomaly-free, restricts the quantum numbers of the fermions in the theory. If we consider the SM fermion content, popular NP scenarios involve a new $U(1)$ gauge group whose fermion quantum numbers are linear combination of the baryon ($B$), lepton ($L$), and lepton flavor ($L_i$) numbers, such as $B-L$, $B - 3L_i$ or $L_i  - L_j$.\footnote{Anomaly-free conditions may require the inclusion of right-handed neutrinos in the theory, especially when a linear combination of $B$ and $L_{(i)}$ is involved.}.

Given that most of the experimental measurements involve first-generation fermions, new gauge interactions with electrons result in either highly suppressed couplings or extremely massive boson. For this reason, we focus on new interactions involving muons and taus in the lepton sector. Following \cite{Foot_1994}, we can define interaction eigenstates as follows:
\begin{eqnarray}
    l_{L,2} &\sim& (1,2)(-1/2,q_{2}^{\prime}), \quad \quad l_{R,2} \sim (1,1)(-1,q_{2}), \\
     l_{L,3} &\sim& (1,2)(-1/2,q_{3}^{\prime}), \quad \quad l_{R,3} \sim (1,1)(-1,q_{3}),
\end{eqnarray}
where the quantum numbers in parenthesis correspond to the representation under $SU(3) \times SU(2) \times U(1) \times U(1)^{\prime}$. Anomaly cancellation imposes the conditions $q_{2}^{\prime} =-q_{3}^{\prime}$, $q_{2} =-q_{3}$, $q_{2}^{\prime} =  q_{2}$. In the following, we will always assume that the $U(1)^{\prime}$ interaction is spontaneously broken, with the corresponding gauge boson acquiring a mass $M_{Z^{\prime}}$.  
The introduction of a new abelian symmetry can induce kinetic mixing between the SM Hypercharge and the new $Z^{\prime}$ already at tree-level, leading to an interaction with electrons once the gauge fields are rotated into the physical states. However, this effect can be entirely avoided at any loop order by imposing the following exact discrete symmetry:
\begin{align}
\ell_{L,2} \leftrightarrow \ell_{L,3}\,, \qquad \ell_{R,2} \leftrightarrow \ell_{R,3}\,, \qquad
B^\mu \leftrightarrow B^\mu\,, \qquad Z^{\prime \mu} \leftrightarrow - Z^{\prime \mu}\,.
\end{align}
We will impose this symmetry for the following models; scenarios where kinetic mixing becomes relevant will be discussed later.
As a first example, we consider the case where $l_{L,2} = \mu_{L}$, $l_{L,3} = \tau_{L}$, $l_{R,2} = \mu_{R}$, $l_{R,3} = \tau_{R}$, where $\mu_{L}$ and $\tau_{L}$ refers to the SM doublets. This is the case of the  spontaneously broken gauged $L_\mu - L_\tau$ symmetry, a model extensively studied in the literature.
The interaction Lagrangian for this model is given by,
\begin{eqnarray} \label{eq:Zpr1}
{\cal L}_{\rm int} = g^\prime Z^\prime_\alpha\left( \bar \mu\gamma^\alpha \mu +  \bar \nu_\mu \gamma^\alpha P_L \nu_{\mu}  -
                \bar \tau\gamma^\alpha \tau -  \bar \nu_\tau \gamma^\alpha P_L \nu_{\tau}\,\right)  , ~~
\end{eqnarray}
From the interaction Lagrangian, we obtain the couplings, $
(g_R)_{ij} = (g_L)_{ij} = \delta_{ij}\,g^\prime\,q^\prime_i$, with $q^\prime_1 = 0$, $q^\prime_2 = 1$ and $q^\prime_3 = -1$. 
The contributions of this model to the dipole Wilson coefficients are given by,
\begin{eqnarray}
\label{eq:CZpr1}
  C^{Z^\prime\,(D)}_{i j}\;= \; {Q_{\ell}} \;( g^\prime\, q^\prime_i)^2\;  \delta_{ij}\;\frac{m_i}{M_{Z^\prime}^2}\;\left(2\, I_3(m_{i}^2/M_{Z^\prime}^2)\,+\, I_4(m_{i}^2/M_{Z^\prime}^2)\right)\,. 
\end{eqnarray}
In this model, the couplings and the Wilson coefficient are flavor-diagonal, and we can only expect contributions to flavor-diagonal dipole moments that scale with the mass of the external state. However, the anomaly-free condition allows for models with flavor-changing couplings, which could induce additional enhancement in dipole transitions. This occurs if the interaction eigenstates do not coincide with the mass eigenstates. In our case, a simple example is obtained if we assign charge $q^\prime_2 = 1$ to the fields $\ell_{L,2} = (\mu_L+\tau_L)/\sqrt{2}$ and $\ell_{R,2} = (\mu_R+\tau_R)/\sqrt{2}$ and $q^\prime_3 = -1$ to $\ell_{L,3} = (\mu_L-\tau_L)/\sqrt{2}$ and $\ell_{R,3} = (\mu_R-\tau_R)/\sqrt{2}$ ~  \cite{Foot_1994,Altmannshofer_2016}.
The interaction Lagrangian in this case is,
\begin{eqnarray} \label{eq:Zpr2}
{\cal L}_{\rm int} = g^\prime Z^\prime_\alpha \left( \,\bar \mu\gamma^\alpha \tau+  \bar \nu_\mu \gamma^\alpha P_L \nu_{\tau}  +
                \bar \tau\gamma^\alpha \mu +  \bar \nu_\tau \gamma^\alpha P_L \nu_{\tau} \,\right)  , ~~
\end{eqnarray}
so that the couplings are now, $
(g_R)_{ii} = (g_L)_{ii} = (g_R)_{1i} = (g_L)_{1i} = (g_R)_{i1} = (g_L)_{i1} = 0$ and $(g_R)_{23} = (g_R)_{32}= (g_L)_{23} = (g_L)_{32}=\,g^\prime$, and the corresponding Wilson coefficients,
\begin{eqnarray}
\label{eq:CZpr2}
  C^{Z^\prime\,(FC)}_{2 2}\;= \; {Q_{\ell}} \;( g^\prime\, )^2\; \frac{m_\mu}{M_{Z^\prime}^2}\;\left(2\, I_3(m_{\tau}^2/M_{Z^\prime}^2)\,+\,\frac{m_\tau}{m_\mu} \, I_4(m_{\tau}^2/M_{Z^\prime}^2)\right)\,, 
\end{eqnarray}
with off-diagonal Wilson coefficients being zero and {the diagonal coefficient $C_{33}$ showing the same parametric structure as Eq.~(\ref{eq:CZpr2}) but with $\tau \leftrightarrow\ \mu$ }.  
More general charge assignments are also possible and would have both flavor diagonal and flavor off-diagonal  couplings even involving the first generation. For example, in the $U(1)_{L_\mu - L_\tau}$ model, we could get general $Z^\prime$ couplings if the lepton mass eigenstates are not interaction eigenstates. This can be obtained in the presence of a doublet scalar field $\phi$ with charge +1 under the new symmetry, with a vev contributing to the lepton mass matrix in the $(1,2)$ and $(1,3)$ entries. In this case, the rotation to the mass basis introduces off-diagonal couplings $ g^\prime Z^\prime_\alpha\, U_{ij}^{L,R} \, \bar f^{i}_{L,R}\gamma^\alpha f^{j}_{L,R} $. Then, the dipole coefficients would read:
\begin{eqnarray}
\label{eq:Zprimeoff}
    C_{i j}&=&{Q_{\ell}} \;(g^{\prime})^{2} \sum_{k=2,3}\left[\frac{(U^{R})_{ik} (U^{R}_{jk})^* \,m_{i} +  (U^{L})_{ik}(U^{L})_{jk}^* \, m_{j}}{ M_{Z^{\prime}}^2 }\; I_3(m_{k}^2/M_{Z^{{\prime}}}^2) \; + \;  \frac{(U^{L})_{ik} (U^{R})_{jk}^*m_{k}}{M_{Z^{\prime}}^2} \;  I_4(m_{k}^2/M_{Z^{\prime}}^2)\right]
\end{eqnarray}
This model could contribute both to flavor conserving and flavor violating observables. In the presence of physical phases, an EDM could also be generated, if left-handed couplings differ from the right-handed ones.

So far, we have considered the addition of new abelian symmetries based on the arguments of anomaly cancellation, assuming that the SM fermions are charged under the new symmetry.
Nonetheless, many models in the literature consider the presence of an additional $U(1)$ dark gauge group, under which the SM particles remain uncharged. This can be employed to generate a portal through kinetic mixing with the SM hypercharge via the Lagrangian term 
\begin{equation}
    \frac{\epsilon}{2} B^{\mu \nu}Z^{\prime}_{\mu \nu}.
\end{equation}
If this additional force is long-range, meaning the gauge boson remains massless, the dark boson does not interact with SM fermions, and therefore, this interaction does not contribute to dipole operators at (one-)loop level. However, if a dark boson mass term is present, the mixing with the photon and the $Z$ boson induces interactions with SM fermions, which are parametrized by the following Lagrangian terms:
\begin{equation}
    \mathcal{L}_{\rm int} \supset e Z^{\prime}_{\mu}\epsilon \left(\cos\theta_W \, J_{em}^{\mu} + \sin \theta_W\, \tan \xi\, J_{Z}^{\mu}\right)
\end{equation}
where $\tan \xi \equiv M^2_{Z^{\prime}}/(M_Z^2-M_{Z^{\prime}}^2)$ and the $Z$ current is defined as $J^\mu_Z \equiv (J^\mu_3 -\sin^2 \theta_W\,J^\mu_{\rm em})/\cos\theta_W$, with $J^\mu_3$ being the current associated with the $T_3$ generator of $SU(2)_L$. This results in the following contributions to the vector couplings:
\begin{eqnarray} 
  \label{eq:Zprime_couplings}
  (g_{L})_{ij}=\, \delta_{ij} \, \frac{e\, \epsilon}{2} \left(\frac{2\cos^2 \theta_W Q_{\ell}^{j}-\left(1+2\sin^2 \theta_W\, Q_{l}^{j}\right)\tan \xi}{\cos \theta_W}\right);\qquad\qquad \quad
  (g_{R})_{ij}= \,\delta_{ij} \, e\, \epsilon \left(\frac{\cos^2 \theta_W Q_{\ell}^{j} -\sin^2 \theta_W\, Q_{l}^{j}\,\tan \xi}{\cos \theta_W}\right).
\end{eqnarray}
The contribution to the dipole coefficients can be easily obtained from Eq.~(\ref{eq:DipGen2}).
Given the fact that this interaction is flavor conserving and real, the dark gauge boson can only contribute to real part of flavor-diagonal coefficient of dipoles.

\subsection{Standard Model Effective Field Theory}\label{sec:SMEFT}

Up to this point, we analyzed NP effects on dipole operators in several specific models. However, if NP is significantly heavier than the electroweak scale, we can integrate out the heavy degrees of freedom to get a tower of higher dimensional operators and treat the SM as an effective field theory (SMEFT). In fact, this is typically needed in most of the theories mentioned in the previous sections to apply the Renormalization Group Evolution (RGE) of the different operators and calculate the observables at energy scales of the order of the lepton masses.
From another perspective, since no NP has been discovered up to the TeV scale, parameterizing short-distance effects using the SMEFT framework turns out to be a powerful tool for making predictions without relying on a specific ultraviolet (UV) theory. 
The SMEFT Lagrangian is parametrized as a tower of $d$-dimensional operators, $O_i^{(d)}$, that are invariant under the SM gauge group:  
\begin{equation}
    \mathcal{L}_{SMEFT} = \mathcal{L}_{SM} + \sum_{d}\sum_{i} \frac{C_{i}^{(d)}}{\Lambda^{d-4}}O_i^{(d)}\,,
\end{equation}
where $C_i^{(d)}$ are the corresponding  Wilson coefficients and $\Lambda$ the ultraviolet cut-off of the effective theory. The SMEFT Lagrangian contains, in principle, an infinite number of operators, whose relevance diminish progressively as $d$ increases. In this review, we focus on the operators that are usually the most relevant, those at dimension six, in the Warsaw basis \cite{Grzadkowski:2010es} \footnote{There is only one possible type of operator at dimension five, the Weinberg operator, which is responsible for the Majorana neutrinos mass.}. Among them, focusing on the lepton sector, only two types of operators can have a tensor current with gauge fields at dimension six, the electroweak dipoles: 
\begin{equation}\label{SMEFT_dipole}
    O_{eB}^{ij} = \bar{l}_{L}^{i} \sigma_{\mu \nu} e_{R}^{j} H B^{\mu \nu}, \qquad\qquad  O_{eW}^{ij} =  \bar{l}_{L}^{i} \sigma_{\mu \nu} \tau^{a} e_{R}^{j} H W^{\mu \nu}_{a},
\end{equation}
where $B^{\mu \nu}$ and $W_{a}^{\mu \nu}$ are the Hypercharge and Weak bosons field strenght tensors, respectively. 
After electroweak symmetry breaking, these operators can be easily matched onto the dimension 5 dipoles in the broken theory,
\begin{equation}\label{eq:SMEFT_LEFT}
    C_{i j} = \frac{8 \pi^2}{e}\frac{v}{\sqrt{2}\,\Lambda^2}\left(\cos\theta_{W}\,C_{eB}^{ij}-\sin\theta_{W}\,C_{eW}^{ij}\right).
\end{equation}
As mentioned above, one of the key roles of SMEFT analysis is the running under RGE, which allows the resummation of large logarithms, and can have a significant impact anytime a sizable coupling or a large separation scale is involved.
Furthermore, the RGE can induce mixing between different operators,
allowing dimension-six operators that are not dipole operators at the UV scale to contribute to dipole-related observables at the low scales of the experiments.

All of these effects are encoded in the $\beta$ function of the Wilson coefficients, which is ruled by the anomalous dimension matrix $\gamma_{ij}$, according to,
\begin{equation}\label{eq:beta_function}
    16 \pi^2 \frac{d\, C_{i}(\mu)}{d\, \mathrm{ ln}\mu} = \gamma_{ji}^{(m)} C_{j}(\mu),
\end{equation}
where $m$ labels the loop order and $\mu$ is the energy scale of the running. Diagonal elements in $\gamma_{ij}$ of Eq.~(\ref{eq:beta_function}), will be responsible for the self-running, which can be sizable if the QCD gauge coupling is involved, while off-diagonal elements mix different operators in the evolution to low energies.

\begin{table}[ht]
    \centering
    \renewcommand{\arraystretch}{1.5}
    \normalsize
    \smallskip

    \begin{subtable}{0.48\linewidth}
        \centering
        \begin{tabular}{>{\centering\arraybackslash}m{2cm} >{\arraybackslash}m{3.5cm}}
            \toprule
            \multicolumn{2}{c}{\textbf{one-loop RGEs: LO}} \\
            \midrule
            $O_{lequ(3)}^{ijlm}$ & $\overline{l}_{L,i}^{\alpha} \sigma_{\mu \nu} e_{R,j} \, \epsilon_{\alpha \beta} \, \overline{q}_{L,l}^{\beta} \sigma_{\mu \nu} u_{R,m}$ \\
            $O_{HW}$ & $H^{\dagger}H W_{\mu \nu}^{a}W_{a}^{\mu \nu}$ \\
            $O_{H\widetilde{W}}$ & $H^{\dagger}H \tilde W_{\mu \nu}^{a}W_{a}^{\mu \nu}$ \\
            $O_{HWB}$ & $H^{\dagger}\tau_a H W_{\mu \nu}^{a}B_{\mu \nu}$ \\
            $O_{H\widetilde{W}B}$ & $H^{\dagger}\tau^aH \tilde W_{\mu \nu}^{a}B_{\mu \nu}$ \\
            $O_{HB}$ & $H^{\dagger}H B_{\mu \nu}B^{\mu \nu}$ \\
            $O_{H\widetilde{B}}$ & $H^{\dagger}H \tilde B_{\mu \nu}B^{\mu \nu}$ \\
            \bottomrule
        \end{tabular}
        \caption{}
        \label{tab:SMEFT_operators_LO}
    \end{subtable}
    \hfill
    \begin{subtable}{0.4\linewidth}
        \centering
        \begin{tabular}{>{\centering\arraybackslash}m{2cm} >{\arraybackslash}m{3.5cm}}
            \toprule
            \multicolumn{2}{c}{\textbf{one-loop RGEs: NLO}} \\
            \midrule
            $O_{ lequ(1)}^{ijlm}$ & $\overline{l}_{L,i}^{\alpha} e_{R,j} \, \epsilon_{\alpha \beta} \, \overline{q}_{L,l}^{\beta} u_{R,m}$ \\
            $O_{eB}^{ij}$ & $\bar{l}^{i}_{L} \sigma_{\mu \nu} e_{R}^{j} H F^{\mu \nu}$ \\
            $O_{eW}^{ij}$ & $\bar{l}^{i}_{L} \sigma_{\mu \nu} \tau^{a} e_{R}^{j} H W^{\mu \nu}_{a}$ \\
            $O_{uB}^{ij}$ & $\bar{l}^{i}_{L} \sigma_{\mu \nu} u_{R}^{j} \tilde{H} F^{\mu \nu}$ \\
            $O_{uW}^{ij}$ & $\bar{l}^{i}_{L}\sigma_{\mu \nu} \tau^{a} u_{R}^{j} \tilde{H} W^{\mu \nu}_{a}$ \\
            $O_{dB}^{ij}$ & $\bar{l}^{i}_{L} \sigma_{\mu \nu} d_{R}^{j} H F^{\mu \nu}$ \\
            $O_{dW}^{ij}$ & $\bar{l}^{i}_{L}\sigma_{\mu \nu} \tau^{a} d_{R}^{j} H W^{\mu \nu}_{a}$ \\
            $O_{W}$ & $\epsilon^{IJK} W_{\mu}^{I \nu} W_{\nu}^{J\rho} W_{\rho}^{K \mu}$ \\
            $O_{\widetilde{W}}$ & $\epsilon^{IJK} W_{\mu}^{I\nu} W_{\nu}^{J\rho} \tilde{W}_{\rho}^{K\mu}$ \\
            \bottomrule
        \end{tabular}
        \caption{}
        \label{tab:SMEFT_operators_NLO}
    \end{subtable}

    \caption{List of operators that mix to the electroweak dipoles at LO and NLO. We define $\tilde{V}^{\mu \nu} = \frac{\epsilon^{\mu \nu \rho\sigma}}{2} V_{\rho\sigma}$, where $V$ refers to either $B$ or $W$, and $\epsilon_{\alpha\beta}$ as the Levi-Civita tensor of $SU(2)$. Operators such as $\mathcal{O}_{eV}$ can induce mixing between dipole operators with different lepton families.}
    \label{tab:SMEFT_operators}
\end{table}

Considering a complete basis for the SMEFT Lagrangian, it is possible to identify all the operators at dimension six that  contribute to the dipole coefficients under RGE.
Neglecting the self-running contribution, the $\beta$ functions at one-loop for the Wilson coefficients of the electroweak dipoles are \cite{Jenkins_2014,Alonso_2014}:
\begin{align}
    16 \pi^{2} \frac{d{C}_{eW}^{ij}}{d\mathrm{ln}\mu} &= -2g_2 N_c C_{lequ(3)}^{ijlm} [Y_u]_{ml} - [Y_e^\dagger]_{ij} \left( g_2 (C_{HW} + i{C}_{H\widetilde{W}}) + g_1 (y_l + y_e) (C_{HWB} + i{C}_{H\widetilde{W}B}) \right)  \label{eq:RGEs_dipolea}\\
     16 \pi^{2} \frac{d{C}_{eB}^{ij}}{d\mathrm{ln}\mu} &= 4g_1 N_c (y_u + y_q) C_{lequ(3)}^{ijlm} [Y_u]_{ml} - [Y_e^\dagger]_{ij} \left( 2g_1 (y_l + y_e) (C_{HB} + i{C}_{H\widetilde{B}}) + \frac{3}{2} g_2 (C_{HWB} + i{C}_{H\widetilde{W}B}) \right)\label{eq:RGEs_dipoleb}\,,
\end{align}
where $g_{1,2}$ are the hypercharge and weak gauge couplings, $Y_{u,e}$ are the up quark and lepton Yukawa matrices and $y_i$ are the SM hypercharges. The indices $i,j,l,m$ label the fermion flavors. The different operators associated to the Wilson coefficients in Eq.~(\ref{eq:RGEs_dipolea},\ref{eq:RGEs_dipoleb}) are shown in Table~(\ref{tab:SMEFT_operators_LO}). 

From these operators, some of them, such as $O_{HV(V)}$ and ${O}_{H\widetilde{V}(V)}$ are Hermitian, requiring their Wilson coefficients to be real. In particular, $O_{HV(V)}$ is CP-even, while ${O}_{H\widetilde{V}(V)}$ is CP odd, so that they contribute to the anomalous magnetic moments and EDMs, respectively. Looking at Eqs. (\ref{eq:RGEs_dipolea}, \ref{eq:RGEs_dipoleb}), LFV in dipole observables through these latter operators could only be triggered by the lepton Yukawa matrix. However, in the SM, $Y_{e}$ is real and diagonal, so that processes such as $\ell_{i}\rightarrow \ell_{j} \gamma$ cannot be induced by these operators at dimension six. Effects induced by a modification of the Yukawa texture are discussed below.

In contrast, the operator $O_{lequ(3)}$ is not hermitian and allows  transitions between different families. As a result, it can affect all the dipole observables, if sources of CP and flavor violation are present in the UV completion of the model. Additionally, the correspondent anomalous dimension is proportional to the up quark Yukawa matrix, so that a large enhancement can occur when top quarks are present in the loop. This operator is particularly interesting because it is generated by integrating out heavy scalars that couple to quarks and leptons in the same interaction, usually referred to as scalar leptoquarks. These particles are predicted by Grand Unified Theory and, in the recent years, have gained popularity as candidates for the explanation of \emph{flavor anomalies} in specific semi-leptonic B meson decays \cite{Becirevic:2018afm} \footnote{As of today, some of these \emph{flavor anomalies} have disappeared due to a recast of the experimental data by the LHCb collaboration. However, some tensions persist when performing a global fit with other experimental measurements. See \cite{Capdevila_2023} for a recent review.}.
\begin{figure}[h]
	\centering
\includegraphics[width=14.5
cm]{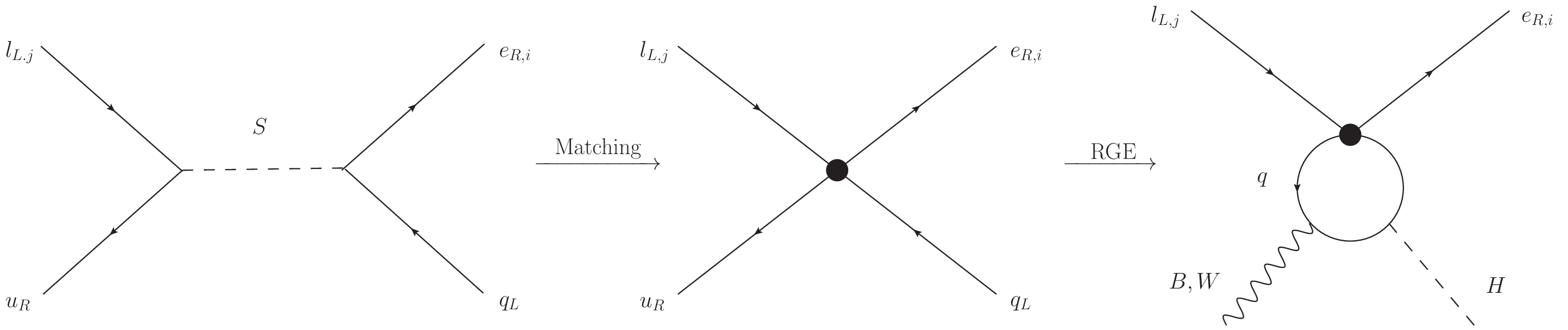}
	\caption{Contribution from a scalar leptoquark to dipole operators, induced via the effective operator  $O_{ lequ(3)}$ after tree-level matching. The dipole is then generated at one-loop through renormalization group evolution (RGE).}
\label{fig:Leptoquark_matching}
\end{figure}
For instance, this operator is generated by a heavy scalar leptoquark with SM quantum numbers (3, 2, 7/6), with interactions
\begin{equation}
    \mathcal{L}_{\rm Lq} \supset - y_{ij}^{qe}\, S \, \overline{q}^{i}_{L} e_{R}^{j} - y_{ij}^{lq}\, \tilde{S} \, \overline{l}^{i}_{L} u_{R}^{j}\, + \,\rm H.c.,
\end{equation}
with $\tilde{S} = i \tau_{2}S^{*}$, as shown in Figure~\ref{fig:Leptoquark_matching}. If we integrate out the heavy scalar field at tree-level by using the equations of motion, the resulting dimension six operator would be, $\overline{l}^{\alpha}_{L,i} u_{R,m}\,\epsilon_{\alpha \beta}\,\overline{q}^{\beta}_{L,l} e_{R,j}$, that is transformed into  $O_{lequ(3)}$ after using Fiertz identities\footnote{Fierz identities are algebraic relations used to rearrange products of gamma matrices and spinors, often applied in SMEFT to rewrite fermion bilinears. Caution is needed in loop calculations, as these identities hold strictly in 4D and may introduce evanescent operators when using dimensional regularization.}. According to Eqs.~(\ref{eq:RGEs_dipolea}, \ref{eq:RGEs_dipoleb}), this operator contributes to dipoles at one-loop level in the RGE by closing the quark loop and attaching one Higgs boson and a B or W gauge field. 

In addition, RGE resummation captures effects beyond the leading log in the one-loop $\beta$ function. These arise from operators that mix with those in Eqs.(\ref{eq:RGEs_dipolea},\ref{eq:RGEs_dipoleb}) and subsequently feed into the dipole. Although suppressed by extra loop factors, large logarithms or unsuppressed SM couplings can result in sizable contributions to dipole observables. A complete list of such operators contributing to the dipole Wilson coefficients at NLO is given in Table~(\ref{tab:SMEFT_operators_NLO}). First-generation dipole observables can be particularly sensitive to scalar four-fermion operators like $O_{lequ(1)}^{ijlm}$ or dipole-type operators $O_{uV}^{ij}$ when top quarks are involved, due to the sizable top Yukawa coupling in the RGE.

Effects induced by finite mixing, namely operators that enter dipoles not in the RGE, are usually dependent on the renormalization scheme or the basis, given the fact that the trace of $\gamma_5$ and the contribution of evanescent operators affect the finite part of the one-loop computation in an ambiguous way. In view of preserving the model independence of the SMEFT analysis, these effects can be taken under control when the two-loop anomalous dimension matrix is considered \cite{Ciuchini:1993ks}.
For an analysis of the two-loop effects in dipole operators see \cite{Panico_2019}. 

As anticipated, relevant dimension six effects can also be triggered by the modification of the Yukawa texture. Within the SMEFT,  this is ruled by operators such as $\mathcal{O}_{f\varphi}^{ij} = H^{\dagger}H \,\overline{f}_{i}\ptwiddle{H}f_{j}$. Indeed, after the electroweak spontaneous symmetry breaking, they modify the SM Yukawa matrices, which are no longer proportional to the fermion
mass matrix. As a result, if $C_{f\varphi}^{ij}$ is complex or off-diagonal, CP or flavor violating interactions with the Higgs and the SM fermions may survive even after rotating the fields in their mass eigenstate basis. As a consequence, this can induce observable effects in EDMs or dipole transitions.

Large effects due to modified Yukawa couplings can stem from the Barr-Zee diagrams in Figure \ref{fig:BarrZee}, with the Higgs boson as the internal scalar. Indeed, in the SM, such contributions are real and flavor diagonal so that they can impact only the anomalous magnetic moment. However, if the NP effects encoded in $O_{e \varphi}$ are not proportional to the SM Yukawa itself, these diagrams feel the modified Higgs interactions in the external lepton line, leading to possible large contribution to EDMs or $\ell_{i} \rightarrow \ell_{j} \ \gamma.$

\section{Dipole observables}\label{sec:Dipobs}

In the previous section, we derived the general expression for the "dipole matrix" at one-loop and discussed additional two-loop contributions that could be relevant. Here, we apply these expressions to the main dipole observables, anomalous magnetic moments, EDMs and LFV processes, and analyze their sensitivity to physics beyond the SM.

\subsection{Muon Anomalous Magnetic Moment}\label{Subsec:muong-2}

As explained in Section~\ref{sec:DipNewPhys}, among all magnetic moments, the muon magnetic moment is generally the most sensitive probe for NP. For a long-time there was a sizable discrepancy between the experimental results and SM predictions that was considered a potential sign of NP. This is mainly due to the fact that the uncertainty in the computation of $a_{\mu}$ is dominated by
the Hadronic Vacuum Polarization (HVP) contribution, which corresponds to the Schwinger’s diagram with the photon propagator dressed
with a hadronic bubble. Due to the fact that the QCD coupling constant becomes large at low energy, this contribution can not be reliably
computed in perturbation theory. For many years, this quantity has been extrapolated from the low-energy $e^{+}e^{-}
\to \rm had.$ data, with the so-called data-driven approach. After the latest result from the Muon g-2 collaboration, combined with the earlier BNL measurement, the current discrepancy with the SM prediction is given by \cite{ParticleDataGroup:2024cfk}:
\begin{equation}
\Delta a_\mu^{\rm exp} \equiv a_\mu^{\rm exp} - a_\mu^{\rm SM} = \left(2.49 \pm 0.48\right) \times 10^{-9}, 
\end{equation}
which corresponds to a 5.2$\sigma$ deviation and confirms the earlier experimental results. For this reason, there is extensive literature exploring possible NP explanations for this discrepancy. Recently, a new lattice determination of the HVP by the BMW collaboration has reduced this discrepancy to the 0.9 $\sigma$-level \cite{Boccaletti:2024guq}. Here, we will discuss the different possibilities in general terms and refer to the comprehensive review articles for a more detailed description of specific models.
Before discussing the different NP models, it is instructive to compare the size of this possible discrepancy with the weak contribution of the SM, which reads
\cite{Fujikawa:1972fe,Aebischer:2021uvt}:
\begin{equation}
a_\mu^{\rm EW} \text{(1 loop)} \simeq \frac{G_F m_\mu^2}{12 \sqrt{2} \pi^2} \left[ 3 -4 \sin^2\theta_W + 8 \sin^4\theta_W \right] \simeq \frac{G_F m_\mu^2}{12 \sqrt{2} \pi^2} \frac{5}{2}= 1.95 \times 10^{-9}\, .
\end{equation}
This means that, with couplings of order of the gauge electroweak couplings and a chirality flip of order of the muon mass, the NP masses should be of the order of $M_w$ to explain this discrepancy. 
In terms of the dipole Wilson coefficient, the electroweak contribution is,
\begin{equation}
\label{eq:EWg-2}
C_{22}^{\rm EW}  =  \frac{5}{2}\, \frac{g^2}{48}\, \frac{m_\mu}{M_W^2} \;\simeq \; \frac{g^2}{20}\, \frac{m_\mu}{M_W^2} \,.
\end{equation}
If we consider the new lattice computation, which agrees with the experimental measurements, the sensitivity to NP is roughly given by the errors in this difference, that if combined quadratically, would give us, $\delta a_\mu^{\rm BMW}  \simeq  0.45 \times 10^{-9}$,
 still of the order of the EW contribution. Given the current uncertainty in the theoretical prediction, the muon g-2 anomaly can be interpreted either as evidence for NP (when considering the data-driven result) or as a sensitivity to heavy NP (when considering the lattice-based result).

From here, we will use the electroweak contributions to determine the sensitivity of the muon anomalous magnetic moment to heavy scales of NP. As an example, we can analyze the contribution of the Type X 2HDM in the decoupling limit, $\cos(\beta -\alpha)\to 0$, and for large $\tan \beta$,
\begin{align}
\label{eq:mu2HDMII}
C_{22}^{\rm 2HDM} &\simeq  \,\frac{ g^2\,m_\mu^3}{{16} \,M_W^2}\,\tan^2 \beta\,\left(  \frac{1}{M_H^2} \,  \left(-\frac{7}{6} - \log \frac{m_\mu^2}{M_H^2}\right) \,+\, \frac{1}{M_A^2} \, \left( \frac{11}{6} + \log \frac{m_\mu^2}{M_A^2} \right)\; - \; \frac{1}{12\,M_{H^+}^2} \right)\; \,,
\end{align} 
which, in the limit $M_H\simeq M_{H^+}\simeq M_A$ reduces to,
\begin{align}
\label{eq:mu3HDMII}
C_{22}^{\rm 2HDM} &\simeq \,\frac{ g^2\,m_\mu^3}{{16} \,M_W^2}\,\frac{\tan^2 \beta}{2\, M_A^2}\,.
\end{align} 
For $M_A$ larger that $M_W$, as we expect in the decoupling limit, we have that this contribution is suppressed by a factor $\, m_\mu^2 \tan ^2\beta/(2 \,M_A^2) \simeq  10^{-6} \tan^2\beta \, M_W^2/M_A^2$, and therefore it is negligible. The situation regarding the muon mass suppression is the same in the non-decoupling limit, $\sin(\beta -\alpha)\to 0$, although in this case, scalars can be lighter than $M_W$, and, as explained in Ref.~\cite{Dedes:2001nx}, could give a sizable contributions to $(g-2)_\mu$ for a scalar mass $\sim 10$~GeV and $\tan \beta \sim 10$.

Even in the Type III 2HDM, the largest contribution, using the Chang-Sher ansatz and with an internal $\tau$ lepton, would be,
\begin{align}
\label{eq:mu2HDMIII}
C_{22}^{\rm 2HDM_{III}} &\simeq \,\frac{ g^2\,m_\mu\,m_\tau^2}{{8} \,M_W^2}\,\lambda_{23}\lambda_{32}\,\left(  \frac{\cos^2(\beta -\alpha)}{M_h^2} \, \left(-\frac{3}{2} - \log \frac{m_\tau^2}{M_h^2}\right)\,+\, \frac{\sin^2(\beta -\alpha)}{M_H^2} \, \left(-\frac{3}{2} - \log \frac{m_\tau^2}{M_H^2}\right)\, {+} \, \frac{1}{\,M_{A}^2}\,  \left(-\frac{3}{2} - \log \frac{m_\tau^2}{M_A^2}\right)\right) \cr =& \frac{ g^2\,m_\mu\,m_\tau^2}{{8} \,M_W^2}\,\lambda_{23}\lambda_{32}\left[\,\left(-\frac{3}{2} - \log \frac{m_\tau^2}{M_h^2}\right)\,\left(  \frac{\cos^2(\beta -\alpha)}{M_h^2} \,+\, \frac{\sin^2(\beta -\alpha)}{M_H^2} \, {+} \, \frac{1}{M_{A}^2}\right)\, -\, \frac{\sin^2(\beta -\alpha)}{M_H^2} \,\log \frac{M_h^2}{M_H^2}\,-\, \frac{1}{M_{A}^2} \,\log \frac{M_h^2}{M_A^2}\right]\,.
\end{align}
 We can make a simple estimate defining an effective mass $1/\tilde M_H^2 = \left( \cos^2(\beta -\alpha)/M_h^2 \,+\, \sin^2(\beta -\alpha)/M_H^2 \, {+} \, 1/M_{A}^2\right)$ and neglecting the "smaller" logarithms, $\log M_h^2/M_{H,A}^2$,
\begin{align}
\label{eq:mu3HDMIII}
C_{22}^{\rm 2HDM} &\simeq \,- \frac{ g^2\,m_\mu m_\tau^2}{{8} \,M_W^2}\,\lambda_{23}\lambda_{32}\,\frac{7}{ \tilde M_H^2}\,.
\end{align} 
where we replaced, $\left(- 3/2 - \log (m_\tau^2/M_h^2)\right) \simeq 7$. Comparing with the EW contribution, Eq.~(\ref{eq:EWg-2}), we see that the type III is suppressed by a factor, $\simeq 10^{-2\,} \,\lambda_{23}\lambda_{32}\,M_W^2/\tilde M_H^2$. Again, this contribution is too small for scalar masses above the EW scale. 

Obviously the smallness of 2HDM contribution is due to the proportionality of Yukawa couplings to the fermion masses. In this situation, in spite of being a two-loop contribution, the Barr-Zee diagram can be relevant. From Eq.~(\ref{eq:2HBarrZee}), assuming that $\rho_{22} \sim m_\mu/v$, and defining an effective loop function, $\tilde f(m_t^2/M_S^2) \sim O(1)$, from the rest of the factors in this equation to make a rough estimate, we have,
\begin{align}
C_{22}^{t} &= {- \frac{3\, \alpha\, Q_{u}^2}{2\pi \,m_t}}\, \frac{g^2\,m_t m_\mu}{4 \,M_W^2}\,  \tilde f(m_t^2/ M_S^2)\,.
\end{align}
Comparing this result with the EW contribution, we see that $C_{22}^{t} /C^{\rm EW}_{22} \simeq 30\,\alpha/(9 \pi) \tilde f(m_t^2/M_S^2) \simeq 10^{-2}\,\tilde f(m_t^2/M_S^2)$, i.e. although relatively small, it is indeed larger that other 2HDM contributions.

The situation can be different in other theories like supersymmetry, where some of this scalar couplings are related to gauge couplings. Using the chargino contribution,
\begin{eqnarray}
\label{eq:muchargino}
  C^{\chi^\pm}_{22}&\simeq& \frac{ g^2\,m_\mu}{M_{\tilde \nu}^2}\;\mu M_2 \tan \beta\; \frac{  I_2(\mu^2/M_{\tilde \nu}^2)\; -  \; I_2(M_2^2/M_{\tilde \nu}^2) }{\mu^2 -  M_2^2}  \,.
\end{eqnarray}
We can make a simple estimate taking $\mu\simeq M_2\simeq M_{\tilde\nu}$ and using  $I_2^\prime(x) \xrightarrow[x\to1]{}{1/16}$. Then, we have,
\begin{eqnarray}
\label{eq:muchargino1}
  C^{\chi^\pm}_{22}&\simeq& \frac{ g^2\,m_\mu}{M_{\tilde \nu}^2}\;  \frac{\tan \beta }{{16}}  \,.
\end{eqnarray}
If we compare with the electroweak contribution, we can see that $C^{\chi^\pm}_{22} \simeq C^{\rm EW}_{22}$ for $M_{\tilde \nu} \simeq {\sqrt{\,\tan \beta}} \, M_W$, i.e. the muon anomalous magnetic moment can be sensitive to $M_{\tilde \nu} \simeq {0.5}$~TeV for  $\tan \beta = 50$. From Eq.~(\ref{eq:neutral1}), we can also check that the neutralino contribution is of the same order. 

The contributions to anomalous magnetic moment from  extra $U(1)$ symmetries with flavor diagonal couplings, given in Eq.~(\ref{eq:CZpr1}) are analogous to the electroweak contribution. Assigning charges $\pm 1$ to the second and third generation leptons, and  for small $m_\mu^2/M_{Z^\prime}^2$, we obtain\footnote{This is a rough estimate. The different numerical factor arises basically from the $g/2$ electroweak coupling and the signs in the different  contributions \cite{Fujikawa:1972fe}.}, $ C^{Z^\prime\,(D)}_{22} / C^{\rm EW}_{22} \simeq {3}\, \left(g'/g\right)^2 (M_W/M_{Z^\prime})^2$ . We can see that this contribution receives no significant enhancement, and, as expected, it is of the same order as the electroweak contribution. This implies that for  $Z^\prime$ couplings of electroweak size, $a_\mu$ would be sensitive to $Z^\prime$ masses around $M_{Z^\prime} \lesssim {100}$~GeV.  In $Z^\prime$ models with flavor-changing coupling, the contribution to the muon anomalous magnetic moment, as given in Eq~(\ref{eq:CZpr2}), can receive important enhancements, proportional to the tau mass. Then, we have, 
$ C^{Z^\prime\,(FC)}_{22} / C^{\rm EW}_{22} \simeq {10}\, m_\tau/m_\mu\, \left(g'/g\right)^2 (M_W/M_{Z^\prime})^2$, and, in this case, for similar size couplings we are sensitive to masses, $M_{Z^\prime} \lesssim {1}$~TeV (for a complete analysis, taking into account other constraints in the model, see Refs.~\cite{Foot_1994,Altmannshofer_2016}).

The same analysis can be repeated in any SM extension, modifying appropriately couplings and masses.  For instance, we can check the contributions from a scalar leptoquark \cite{Bauer:2015knc,Popov:2016fzr,Crivellin:2017zlb}. The enhancement in minimal models is provided by the top quark mass, and we would have, $ C^{\rm LQ}_{22} / C^{\rm EW}_{22} \sim m_t/m_\mu\, \lambda_L \lambda_R/g^2 (M_W/M_{\rm LQ})^2$, with $\lambda_{L,R}$ the leptoquark couplings. 
From here, assuming couplings of order $g$, the muon anomalous magnetic moment would be sensitive to scalar leptoquark masses as large as $M_{\rm LQ} \lesssim 3.3$~TeV. However, in this case, we can not forget the contribution to the muon mass itself \cite{Czarnecki:2001pv}. From Eq.~(\ref{eq:masscorrection}), we would have a correction to the mass, $\Delta m_\mu / m_\mu \simeq m_t/m_\mu \; \lambda_L \lambda_R/(16 \pi^2) \simeq 0.46$, for electroweak size couplings, $\lambda_L\simeq \lambda_R\simeq g$.

Finally, the sensitivity of the anomalous magnetic moment to NP can be studied in a model-independent way using the SMEFT framework. Since the same SMEFT dipole operators contribute to all the dipole observables discussed in this review, the results are largely analogous across them. The SMEFT parametrization provides a general and model-independent estimate of the energy scale to which $\Delta a_{\mu}$ is sensitive. Taking the electroweak contribution as a reference and using Eqs.(\ref{eq:advsC}, \ref{eq:SMEFT_LEFT}), we estimate:
\begin{equation}
\label{eq:muong-2_senstitivity} 
\frac{\mathrm{Re}[C_{eB}^{22}],\mathrm{Re}[C_{eW}^{22}]}{\Lambda^2} \, \sim \, 10^{-5} \; 
\rm TeV^{-2} 
\end{equation} 
Running and mixing effects depend on the matching scale to the UV theory and the energy range of the RGE evolution, making it difficult to provide a fully general result. However, in the multi-TeV regime, Ref.~\cite{Aebischer:2021uvt} shows that $\Delta a_{\mu}$ is primarily sensitive to the tensor operator $C_{lequ(3)}$ involving the top and charm quarks, in addition to the SMEFT dipole operators themselves.

\subsection{Electric Dipole Moments}

Electric dipole moments (EDMs) of the electron, neutron, or nuclei are highly suppressed in the SM due to its limited sources of CP violation. However, according to Sakharov condition, CP-violation is one of the key ingredients required to explain the observed dominance of matter over anti-matter in the universe, through a mechanism known as baryogenesis \cite{Sakharov:1967dj}. It is well known that the amount of CP-violation in the SM it is not enough for a successful baryogenesis \cite{doi:10.1142/S0217732394000629,PhysRevD.51.379}, requiring the presence of new sources of CP-violation coming from theory beyond the SM.
For this reason, extensions of the SM typically introduce additional sources of CP violation, making EDMs highly promising probes for NP.

From Eq.~(\ref{eq:effL}), we have that EDMs are given by
$d_\ell = -e/(4 \pi^2) \; \text{Im} (C_{\ell \ell})$. Therefore we need the diagonal $C_{\ell \ell}$ Wilson coefficient to have an imaginary part, in the basis where charged lepton masses are real and diagonal. From Eqs.~(\ref{eq:DipGen1}- \ref{eq:DipGen4}), we can see that at one-loop, only the terms with internal chirality change and complex $g_L,g_R$, or $y_{R},y_{L}$,  can give an imaginary part.

As seen in section \ref{sec:2HDM}, in 2HDMs, this is only possible at one-loop through a mixing scalar-psedoscalar. 
At one-loop, in the type X 2HDM, in the decoupling limit, $\cos (\beta-\alpha) \to 0$, where $M_{h_2}\simeq M_{h_3} \simeq M_A$ \footnote{Here, $M_A$ refers to the heavy masses in this limit, that would correspond to the pseudoscalar mass in the CP conserving limit}, using Eq.~(\ref{eq:DipGen1}) and Eq.~(\ref{eq:2HDMCeps}), we obtain,
\begin{align}
\text{Im} \left\{ C_{ii}^{\rm 2HDM} \right\} = \sum_a\frac{m_i Q_i \,\text{Im}  \left\{ (y_R^a)_{ii} (y_L^a)_{ii}^* \right\}}{M_{h_a}^2} I_2 (x_{i a})  \,=\,  \frac{g^2 m_i^3 Q_i}{{4} M_W^2} \epsilon \tan^2  \beta\, \left(\frac{I_2(x_{i2})}{M_{h_2}^2} + \frac{I_2(x_{i3})}{M_{h_3}^2}\right) \simeq -\frac{g^2 m_i^3}{{32} M_W^2} \frac{\epsilon \tan^2  \beta}{M_A^2}\, \left(3 + 2 \log (m_i^2/M_{A}^2)  \right)\,. 
\end{align}
As expected, this contribution is suppressed by the lepton mass cube and this makes it too small. Numerically, for the electron, we have,
\begin{align}
d_e  = - \frac{e}{4 \pi^2}\, \text{Im} \left\{ C_{11}^{\rm 2HDM} \right\} \simeq {4 \times10^{-22} \, \epsilon\, \frac{M_W^2\,\tan ^2\beta }{M_A^2}\, \left(1 - \frac{\log(M_W^2/M_A^2)}{25}\right) ~\text{e $\cdot$ MeV}^{-1} \simeq   10^{-32} \, \epsilon\, \frac{M_W^2\, \tan^2 \beta }{M_A^2}\, \left(1 - \frac{\log(M_W^2/M_A^2)}{25}\right) ~\text{e $\cdot$ cm}}
\end{align}
that, even if we assume $\epsilon \sim {\cal O}(1)$ is too small for the present experimental sensitivity. In type III 2HDM, using the Cheng-Sher ansatz, we would increase this contribution by a factor $m_\tau/m_e \simeq 3.5 \times 10^3$, assuming $ \lambda_{ij}\sim {\cal O}(1)$. In these conditions, we would obtain a bound {$\epsilon \lesssim 10^{-1}$} on the CP violating mixing.

However, as for the anomalous magnetic moment, the main contribution in 2HDM comes from the Barr-Zee diagram. From Eq.~(\ref{eq:2HBarrZeeCP}), in the decoupling limit, we have,
\begin{align} d_e = - \frac{e}{4 \pi^2}\,
\text{Im} \left\{ C_{11}^{\rm 2HBZ} \right\} \simeq  \frac{3\, e\,\alpha\, Q_{u}^2}{{16} \pi^2}\, \frac{g}{M_W}\,\epsilon \frac{\rho_{11}}{\sqrt{2}} \left( 1 {-} \frac{m_t^2}{2 M_{h_3}^2} \log^2\frac{m_t^2}{ M_{h_3}^2} \right) \simeq {1 \times 10^{-26}}\, \epsilon \, \left( 1 {-} \frac{m_t^2}{M_{h_3}^2} \log^2 \frac{m_t^2}{2 M_{h_3}^2}\right) ~\text{e $\cdot$ cm} \,,
\end{align}
where we assumed $\rho_{11}/\sqrt{2}\simeq m_e/v$, $\xi_{i}^{u}\simeq 1$ and used   $\lim_{x_{t2}\to 0} f(x_{t2})\simeq \lim_{x_{t3}\to 0} f(x_{t3}) \simeq \frac{1}{2} x_{t3}\log^2 (x_{t3})$ and $f(x_{t1})\simeq 1$ and slowly varying around $x_{t1}\simeq 1$. 
Therefore, the Barr–Zee diagram dominates the one-loop contribution to the electron EDM by approximately six orders of magnitude, giving a sizeable enhancement. As a result, this rough estimate suggests that scalar–pseudoscalar mixing,  $\epsilon$, would be constrained to be $\epsilon \lesssim { 10^{-4}}$ by the present upper bound on $\vert d_{e}\vert$, considering only the top-quark loop. For a more detailed analysis within 2HDMs, see Ref.~\cite{Altmannshofer_2020}.

As we saw in the previous section, supersymmetric models can generate larger contribution, not suppressed by additional Yukawa couplings. Indeed, assuming an ${\cal O}(1)$ phase in the $\mu$ parameter, appearing both in the chargino, Eq.~(\ref{eq:charginomat}) and the neutralino, Eq.~(\ref{eq:neutralmatrix}), mass matrices, leads to the well-known Supersymmetric CP problem \cite{Dugan:1984qf,Masiero:2001ep}. 

If we consider the chargino contribution, from Eq.~(\ref{eq:chargino3})  we obtain,
\begin{align} d_e = - \frac{e}{4 \pi^2}\,
\text{Im} \left\{ C_{11}^{\chi^+} \right\} \simeq{-}
 \frac{e\, g^2\,m_e}{4\pi^2\,M_{\tilde \nu}^2}\;\text{Im} \left\{M_2\,\mu \right\}\,\tan \beta\; \frac{I_2(\mu^2/M_{\tilde \nu}^2)\, -\,I_2(M_2^2/M_{\tilde \nu}^2) }{\mu^2 -  M_2^2}   \,,
\end{align}
which, in the limit $\mu\simeq M_2\simeq M_{\tilde\nu}$ and using  $I_2^\prime(x) \xrightarrow[x\to1]{} {1/16}$, gives
\begin{align} d_e \simeq
 \frac{- e\,g^2\,m_e}{{64} \pi^2\,M_{\tilde \nu}^2}\;\frac{\text{Im} \left\{M_2\,\mu \right\}}{M_{\tilde \nu}^2}\,\tan \beta\;  \simeq \,  10^{-24} \,\frac{M_W^2}{M_{\tilde \nu}^2}\, \sin \phi_\mu \,\tan \beta\;~ \text{e $\cdot$cm}\,.
\end{align}
where $\phi_\mu$ is the relative phase between $M_2$ and $\mu$ in the basis {of} diagonal and real charged-lepton masses. As a result, if we take $\sin \phi_\mu \sim {\cal O} (1)$, the present electron EDM limit implies that the supersymmetric masses, $M_{\tilde \nu}$, must be larger than $\sim$ {40}$\sqrt{\tan \beta}$~TeV, well above the electroweak scale. Alternatively, if we lower the sneutrino mass to $M_{\tilde \nu}\simeq 1$~TeV, we can an upper bound on the relative phase $\sin \phi_\mu \lesssim {3.5 \times 10^{-5}/\tan \beta}$. 

Clearly, these very strong constraints tell us that this $\mu$ parameter, must be (very approximately) real for masses near the electroweak scale. Indeed, this is a serious possibility, if we remember that the only observed phase  in the Cabibbo-Kobayashi-Maskawa sector is associated to the flavor sector. Then, even assuming $\mu$ real, there are other phases in the MSSM that could contribute to EDMs \cite{Ross:2004qn,Calibbi:2008qt,Hisano:2008hn,Calibbi:2009ja}.

Regarding contributions to EDM from extra $U(1)$ symmetries, given that flavor-diagonal gauge couplings are real, we must consider models with flavor off-diagonal couplings. The corresponding Wilson coefficient, given in Eq~(\ref{eq:Zprimeoff}), is
\begin{eqnarray}
   d_e = - \frac{e}{4 \pi^2}\,
\text{Im} \left\{ C_{11}^{Z^\prime} \right\} \; \simeq \;-\frac{e\,(g^{\prime})^{2}}{4 \pi^2}\, \sum_{k=2,3} \frac{  m_{k}}{{2}M_{Z^{\prime}}^2} \, \text{Im} \left\{ (U^{L})_{1k} (U^{R})_{1k}^*\right\} \,
\end{eqnarray}
where we assumed $M_{Z^{\prime}} \gg m_{k}$ and only the internal chirality changing contribution can be complex if right-handed and left-handed mixings are complex and different. In this expression there is an apparent enhancement, due the tau mass over the electron mass, but, as we saw explicitly in the supersymmetry case, in section \ref{sec:SUSYCij}, the smallness of the mixings can partially compensate this enhancement. Indeed, as shown in section \ref{sec:Zprime}, these off-diagonal couplings can come from the diagonalization of the Yukawa matrix in the presence extra scalars with non-vanishing $L_\mu - L_\tau$ charge. As a result, these couplings contribute to different flavor changing observables so that can be potentially strongly constrained. Therefore, they are expected to be small in full models. In addition to this presence of observable phases is also not trivial \cite{Buras:2021btx}.

For the SMEFT analysis, the scenario is similar to the anomalous magnetic moment, with the difference that the NP scale that can be probed by the electron's EDM is much higher, due to the better sensitivity of the experiments. Indeed, with the same argument, we can estimate
\begin{equation}
    \frac{\mathrm{Im}[C_{eB}^{22}],\mathrm{Im}[C_{eW}^{22}]}{\Lambda^2} \, \sim  \, 10^{-12} \; \rm TeV^{-2},
\end{equation}
which is seven order of magnitude larger than Eq.~(\ref{eq:muong-2_senstitivity}). As a result, from Ref.~\cite{Ardu:2025rqy, Kley:2021yhn}, a larger class of operators is relevant, even if generated far above the TeV scale. 
In particular, beyond the tensor operators, sizable contributions may arise from Barr–Zee two-loop diagrams involving $C_{eH}$, as well as from NLO effects in the one-loop RGE of $C_{lequ(1)}$ with the top quark. 

For a top-down analysis of potential NP contributions to EDMs in the context of scalar leptoquarks matched onto SMEFT operators, we refer the reader to \cite{Dekens_2019}.

For a review of EDMs contributions in these and other models, see Ref.~\cite{Cesarotti:2018huy}.

\subsection{Flavor-Changing Dipole Transitions}

Flavor-changing dipole processes are among the most sensitive probes of physics beyond the SM. In the leptonic sector, the branching ratio BR($\mu \to e \gamma$)  is among the most sensitive experimental observables for flavor-changing processes\footnote{BR($\mu \to e e e$) and $\mu$--$e$ conversion in nuclei are expected to reach comparable sensitivity in the near future \cite{Mu3e:2020gyw,Bernstein:2019fyh,COMET:2018auw}}. Indeed, after the discovery of neutrino oscillations, lepton flavor-violating transitions are expected in the SM with massive neutrinos. However, these contributions are extremely suppressed, well beyond the reach of current experiments, making any observed signal a clear indication of NP.

The analysis of different extensions of the SM that contribute to LFV follows the same strategy used in the previous sections. However, unlike the anomalous magnetic moments and EDMs, BR($\mu \to e \gamma$) is a directly measurable physical observable and, as such, is independent of the chosen basis. Nevertheless, it is often convenient to work in the basis where the charged lepton Yukawa matrix is diagonal and real, in which case, from  Eq.~(\ref{eq:BRmueg}), this observable is proportional to $|C_{21}|^2 + |C_{12}|^2$. Therefore, the main feature of extensions of the SM that contribute to BR($\mu \to e \gamma$) is the presence of additional flavor structures beyond the SM Yukawa matrices that are not simultaneously diagonalizable, giving rise to lepton flavor changing interactions.  

Therefore, neglecting left-handed neutrino masses, in type I, II, X and Y 2HDM, all of which include a $Z_2$ symmetry that restricts each fermion species to couple to a single Higgs doublet, there are no contributions to $\mu \to e \gamma$. In type III 2HDM, and in the decoupling limit, from Eq.~(\ref{eq:W2HDMIII}), we have,
\begin{align}
\label{eq:meg2HDMIII}
C_{2 1} = C_{1 2}^* =\, {-} &m_{\mu}\, \rho_{k2}^* \rho_{k1}\left( \frac{I_1(m_k^2/M_H^2)}{2 M_H^2 }  \,+\,  \frac{I_1(m_k^2/M_A^2)}{2 M_A^2} \,+\, \frac{J_1(m_\nu^2/M_{H^+}^2)}{M_{H^+}^2} \right)\;{-} 
\; m_k\,\rho_{2k} \rho_{k1}  \left(\frac{I_2(m_k^2/M_H^2)}{2 M_H^2 } \,+\, \frac{I_2(m_k^2/M_A^2)}{2 M_A^2} \right) \cr
\simeq &\, m_{\mu}\, \left(\rho_{12}^*\, \rho_{11} \,+\,\rho_{22}^* \,\rho_{21}  \,+\,\rho_{32}^* \,\rho_{31} \right) \frac{1}{{48} M_{A}^2}\; - 
\; {(m_{\tau}\,\rho_{23}\, \rho_{31} + m_{\mu}\,\rho_{22}\, \rho_{21})}  \frac{1}{{4}M_A^2}\,\left(\frac{3}{2} + \log \frac{m_\tau^2}{M_A^2} \right)\,.
\end{align} 
If we apply the Chang-Sher ansatz, this results in
\begin{align}
\label{eq:meg2HDMIIICS}
C_{2 1} 
\simeq &\, m_\mu\,\lambda_{32}^*\,\lambda_{31} {g}  \frac{m_\tau \sqrt{m_e m_\mu}}{{96}\,M_W^2\, M_{A}^2}\; - 
\; m_\tau\,\lambda_{23}\, \lambda_{31}{g}  \frac{m_\tau \sqrt{m_e m_\mu}}{{8}\, M_W^2\,M_A^2}\,\left(\frac{3}{2} + \log \frac{m_\tau^2}{M_A^2} \right)\,,
\end{align} 
and the branching ratio would be,
 \begin{eqnarray}
\label{eq:BRmueg2H}
{\rm BR}(\mu\to e\gamma) \simeq { \cfrac{3\,\alpha }{4\pi}\, \frac{m_e}{m_\mu} \frac{m_\tau^4}{M_A^4} \, \lambda_{23}^2\, \lambda_{31}^2\, \left(\frac{3}{2} + \log \frac{m_\tau^2}{M_A^2}\right)^2\,\simeq 5.0 \times 10^{-11} \, \frac{M_W^4}{M_A^4} \, \lambda_{23}^2\, \lambda_{31}^2\, \left(1 - \frac{1}{6}\,\log \frac{M_W^2}{M_A^2}\right)^2}
\end{eqnarray}

From here, the present bound, BR($\mu \to e \gamma$) $\leq 3.1 \times 10^{-13}$, would require $M_A \gtrsim 300$~GeV for $\lambda_{ij} \sim {\cal O} (1)$. Even in this case, the Barr-Zee diagram can provide an enhanced contribution. As the discussion is similar to that of the previous section, we refer to \cite{Chang:1993kw,Davidson:2016utf} for further details on this contribution.

In supersymmetric models, slepton mass matrices provide additional flavor structures independent of the Yukawa matrices and we can expect contributions to the $\mu \to e \gamma$ process. If consider a general charged slepton mass matrix, the neutralino contribution to the $C_{21}$ Wilson coefficient is,

\begin{align}
\label{eq:neutralmeg1}
C^{\chi^0}_{{12}}/Q_{\tilde{l}}&\simeq {\frac{g^2}{2} \,\frac{ m_\mu\,\mu (M_{\tilde l_L}^2)_{12}\tan \beta }{M_{\tilde l_L}^4} \,
 \; \, \left( \,M_1 t_W^2\,\left(    \; \frac{ J_{2}(\mu^2/M_{\tilde l_L}^2) -J_{2}(M_1^2/M_{\tilde l_L}^2) }{ \mu^2 - M_1^2 }-\frac{- \mu^{2}J_{2}^{\prime}(\mu^2/M_{\tilde l_L}^2) +M_{1}^{2}J_{2}^{\prime}(M_1^2/M_{\tilde l_L}^2) }{ M_{\tilde l_L}^2(\mu^2 - M_1^2) }\right) \,  - \, \frac{1}{t_{W}^{2}} (M_1 \leftrightarrow M_2) \right)}\,, 
 \end{align} 
 while, 
 \begin{align}
 \label{eq:neutralmeg2}
C^{\chi^0}_{{
21}}&\simeq {-g^2 \,\frac{ m_\mu\,\mu \, (M_{\tilde l_R}^2)_{21}\tan \beta }{M_{\tilde l_R}^4} \,
 \; M_1 t_W^2\,\left( \; \frac{ J_{2}(\mu^2/M_{\tilde l_R}^2) -J_{2}(M_1^2/M_{\tilde l_R}^2) }{ \mu^2 - M_1^2 }-\frac{- \mu^{2}J_{2}^{\prime}(\mu^2/M_{\tilde l_R}^2) +M_{1}^{2}J_{2}^{\prime}(M_1^2/M_{\tilde l_R}^2) }{ M_{\tilde l_R}^2(\mu^2 - M_1^2) }\right)} \,.
 \end{align}

 We can make a simple estimate taking the limit of equal masses, $\mu\simeq M_1 \simeq M_2\simeq M_{\tilde l_{L,R}}$. Then we get,
\begin{align}
\label{eq:neutralmeg3}
C^{\chi^0}_{{12}}/Q_{\tilde{l}}&\simeq { \frac{g^2}{120} \, m_\mu\,\frac{ M_2\,\mu \tan \beta }{M_{\tilde l_L}^4} \,\frac{(M_{\tilde l_L}^2)_{12}}{M_{\tilde l_L}^2}
 \; \, \left( 1- \, \frac{t_W^2 M_1}{M_2}\,\right)}\cr
 C^{\chi^0}_{21}/Q_{\tilde{l}}&\simeq { \frac{g^2}{60} \,m_\mu\, \frac{ M_1\,t_{W}^{2}\,\mu \tan \beta }{M_{\tilde l_R}^4} \,\frac{(M_{\tilde l_R}^2)_{21}}{M_{\tilde l_R}^2}} \,. 
 \end{align} 
The branching ratio is then \cite{Ciuchini:2007ha},
\begin{eqnarray}
\label{eq:BRmuegMSSM}
{\rm BR}(\mu\to e \gamma) & =  \cfrac{3\,\alpha_{\rm em} }{\pi\,G_F^2\, m_\mu^2}\, \left(|C_{2 1}|^2+|C_{1 2}|^2\right) \simeq {\cfrac{8\,\alpha_{\rm em} }{4\times 10^3\,\pi}\, \cfrac{M_W^4\,\mu^2 \,M_2^2\,\tan^2 \beta}{M_{\tilde l}^8} \left( \left|(\delta^\ell_L)_{{12}}\right|^2 \left( 1- \,\cfrac{M_1 t_W^2}{ M_{2}^2}\,\right)^2 +  \left|(\delta^\ell_R)_{21}\right|^2 4\, \cfrac{M_1^2 t_W^4}{M_2^2}\right)} \cr &  {
\simeq 3 \times 10^{-7} \,\left(\cfrac{500 ~\text{GeV}}{M_{\tilde l}}\right)^4 \,  \left(\cfrac{\tan  \beta}{5}\right)^2\, \left( \left|(\delta^\ell_L)_{12}\right|^2 \left( 1- t_W^2 \right)^2 + \left|(\delta^\ell_R)_{21}\right|^2 \, 4\, t_W^4\right)}\,,  
\end{eqnarray}
where  we defined the MI as, $(\delta^\ell_{L,R})_{ij} = (M_{\tilde l_L}^2)_{ij}/M_{\tilde l_L}^2$, and took all supersymmetric masses equal to $M_{\tilde l}$. The present bound on $\text{BR}(\mu \rightarrow e \gamma) < 3.1 \times 10^{-13}$, implies that for $M_{\tilde l} = 500$~GeV and $\tan \beta = 5$, we have $|(\delta^\ell_{L})_{12}|,|(\delta^\ell_{R})_{21}| \lesssim {10^{-3}}$ . This means that the flavor structure in the slepton mass matrices can not be arbitrary and must be related with the flavor structure in the Yukawa matrices, perhaps by a flavor symmetry \cite{Ross:2004qn, Calibbi:2009ja}.

Contributions to $\mu \to e \gamma$ from $U(1)^\prime$ models are only possible in models with general flavor off-diagonal couplings. The Wilson coefficient is given in Eq~(\ref{eq:Zprimeoff}). These contributions would be proportional to the off-diagonal $Z^\prime$ couplings, times a fermion mass. For instance, taking the internal chirality changing contribution, that could be enhanced by the tau mass, we would have,
\begin{eqnarray}
\label{eq:BRmuegZpr}
{\rm BR}(\mu\to e \gamma)  = & \,\cfrac{3\,\alpha_{\rm em} }{\pi\,G_F^2\, m_\mu^2}\, (g^\prime)^4 \cfrac{m_\tau^2}{{4}M_{Z^\prime}^4}\, \ \left(|(U^L)_{23} (U^R)^*_{13}|^2+|(U^R)_{23} (U^L)^*_{13}|^2\right) \qquad\cr \simeq &\, {2}\times 10^4 \left(\cfrac{g^\prime}{g}\right)^4\left(\cfrac{M_W}{M_{Z^\prime}}\right)^4 \left(|(U^L)_{23} (U^R)^*_{13}|^2+|(U^R)_{23} (U^L)^*_{13}|^2\right)\, .
\end{eqnarray}
Only from this process, assuming $g^\prime \simeq g$, we would obtain, $|(U^L)_{23} (U^R)^*_{13}|,|(U^R)_{23} (U^L)^*_{13}|  \leq 10^{-9} (M_{Z^\prime}^2/M_W^2)$.

 Nevertheless, at this level, these off-diagonal are arbitrary and could only be defined in a complete model. In the example shown in Section \ref{sec:Zprime}, these off-diagonal  couplings are determined by the diagonalization of the Yukawa matrix in the presence extra scalars with non-vanishing $L_\mu - L_\tau$ charge and $(U^{L,R})_{23}$ or $ (U^{L,R})^*_{13}$, are expected to be small. 
 
Finally,the SMEFT approach allows us to estimate the sensitivity to the NP scale from the measurement of ${\rm BR}(\mu\to e \gamma)$:
\begin{equation}
    \frac{\vert C_{eB}^{21(12)}\vert,\vert C_{eW}^{21(12)}\vert}{\Lambda^2} \, \sim  \, 10^{-10} \; \rm TeV^{-2}\,.
\end{equation}
Therefore, when considering the first generations, the sensitivity to the off-diagonal dipole operators is nearly of the same order as to the imaginary part of the diagonal operator involving the first generation. Since the RGE is identical for all dipole operators, we will not discuss them further. It is, however, interesting to explore possible connections between the various observables, as they can be related by a flavor rotation. Indeed, if the NP allows for LFV, a misalignment between flavor and mass eigenstates can arise, leading to correlations among different observables.
If we assume that the muon $g$–2 sensitivity is saturated by NP, or, seen in another way, that the discrepancy between the data-driven theoretical prediction and experiments is entirely due to NP, the present bound on $\mathrm{BR}(\mu \rightarrow e \gamma)$ is such that \cite{Isidori:2021gqe}
\begin{equation}
    \vert \epsilon_{12(21)} \vert \lesssim 10^{-5};\quad \quad \epsilon_{12(21)} = \frac{C_{eB}^{12(21)}}{\mathrm{Re}[C_{eB}^{22}]} ,\frac{C_{eW}^{12(21)}}{\mathrm{Re}[C_{eW}^{22}]}
\end{equation}
where $\epsilon_{ij}$ parametrize the flavor alignment.

\section{Conclusions}
\label{sec:conclusions}

The SM of particle physics has been tested with exceptional accuracy across a wide range of energy scales, becoming a well established theory for the description of fundamental interactions. However, it fails to explain several phenomena such as the nature of Dark Matter, the neutrino oscillations and the dominance of the matter of the antimatter. While the physics behind these processes may be at energy scale far beyond the reach of the present colliders, its effect could be indirectly detected in low energy experiments. Given the impressive sensitivity achieved in precision physics experiments, along with the expected improvements, it is worthwhile to explore how NP impacts one of the best tested observables: the dipole moments.

In this review, we presented a pedagogical analysis of the sensitivity of dipole transitions to NP beyond the SM, pointing out their intimate connection with the masses of SM fermions. We discussed why dipole-related observables are, and will likely remain in the foreseeable future, among the most sensitive probes of SM extensions. 

As we have shown throughout this work, they offer insights into phenomena such as Lepton Flavor Violation, CP violation, and heavy particle dynamics, remaining at the forefront of efforts to uncover physics beyond the SM.

We provided general one-loop expressions for all dipole observables in terms of generic couplings, applicable to a wide range of models. Additionally, we examined the Barr-Zee two-loop contributions in the presence of new scalar interactions.  These general results were then applied to a selected set of SM extensions, including two Higgs doublet models, supersymmetric frameworks, models with extra $U(1)$ symmetries, and SMEFT, to analyze the sensitivity of dipole observables to the masses and parameters of these theories. 

In particular, for each of the three classes of observables studied, anomalous magnetic moments, electric dipole moments and Lepton Flavor Violation transitions, we focused on a particular generation, i.e. the one that we consider the most promising for the search of NP. Regarding the g-factor, we highlighted why the anomalous magnetic moment of the muon deserves special attention, since it is the most sensitive to NP and still presents some discrepancies between theory and experiments. For the EDMs, we focused on the electron one, whose upper bound is several orders of magnitude stronger than the other generations, and it is the most sensitive observable to NP. Finally, concerning Lepton Flavor Violation, we investigated the effects of beyond the SM physics on the $\mu \to e \gamma$ transition, given its much stronger upper bounds than the other flavor transitions. In fact, it is currently the only flavor-violating observable that is being directly searched for.

This work is intended only as an introductory review for readers arriving at the field for the first time, offering a selection of what we consider the most important concepts in the subject. It is meant to serve as a starting point, and we encourage interested readers to explore the extensive literature and existing reviews for a more in-depth understanding.

 \textit{}

\begin{ack}[Acknowledgments]%
We would like to thank our collaborators on the topics discussed in this review,  from whom we have learned much of what we present here,  especially A. Masiero, L. Silvestrini, P. Paradisi, L. Calibbi, G.G. Ross, J. Jones-Pérez, M. Ardu, G. Barenboim, F.J. Botella, and many others.
We acknowledge financial support from the Spanish  Grant PID2023-151418NB-I00 funded by MCIU/AEI/10.13039/501100011033/ FEDER, UE and from Generalitat Valenciana projects CIPROM/2021/054 and CIPROM/2022/66.
\end{ack}


\bibliographystyle{unsrt} 
\bibliography{reference}

\end{document}